\documentclass[11pt,a4paper]{article}
\usepackage{authblk}
\usepackage{lmodern}
\usepackage[T1]{fontenc}
\usepackage[utf8]{inputenc}
\usepackage[english]{babel}
\usepackage{amsmath}
\usepackage{amsfonts}
\usepackage{amssymb}
\usepackage{graphicx}
\usepackage{kpfonts}
\usepackage{booktabs}
\usepackage{float}
\usepackage{cite}
\usepackage{tikz}
\usepackage{lscape}
\usepackage{caption}
\usepackage{booktabs}

\usepackage{subcaption}
\usepackage{longtable}
\usepackage[multiple]{footmisc}
\usepackage{tabularx}
\newcolumntype{L}{>{\raggedright\arraybackslash}X}
\usepackage[font=small,labelfont=bf,margin=\parindent,tableposition=top]{caption}

\usepackage{array}
\newcolumntype{N}{>{\centering\arraybackslash}m{.5in}}
\newcolumntype{G}{>{\centering\arraybackslash}m{2in}}

\usepackage{dsfont}
\usepackage[left=2cm,right=2cm,top=2cm,bottom=2cm]{geometry}
\usepackage{scalerel,stackengine}
\stackMath
\newcommand\reallywidehat[1]{%
\savestack{\tmpbox}{\stretchto{%
  \scaleto{%
    \scalerel*[\widthof{\ensuremath{#1}}]{\kern-.6pt\bigwedge\kern-.6pt}%
    {\rule[-\textheight/2]{1ex}{\textheight}}
  }{\textheight}%
}{0.5ex}}%
\stackon[1pt]{#1}{\tmpbox}%
}
\DeclareMathOperator*{\argmin}{arg\,min}
\newcommand\norm[1]{\left\lVert#1\right\rVert}

\title{Sustainable Investing and the Cross-Section of Returns and Maximum Drawdown}
\author[1,3,4]{Lisa R. Goldberg}
\author[1,2]{Saad Mouti\thanks{Corresponding author.\\ Email addresses: {lrg@berkeley.edu} (L. R. Goldberg), {smouti@pstat.ucsb.edu} (S. Mouti).}}

\affil[1]{Department of Economics, UC Berkeley}
\affil[2]{Consortium for Data Analytics in Risk, UC Berkeley}
\affil[2]{Department of Statistics and Applied Probability, UC Santa Barbara}
\affil[4]{Aperio by BlackRock, Sausalito\vspace{1cm}}

\date{%
{\parbox{\linewidth}{\centering%
  October 27, 2022\thanks{We are thankful for Robert Anderson, Wachi Bandara, Jeff Bohn, Claudia Bolli, Dangxing Chen, Peter Clark, Xiaowu Dai, Jeremy Evnine, Tingyue Gan, Nicholas L. Gunther, Jim Hawkey, Kathleen Houssels, Paul Jung, Ran Leshem, Raymond Leung, Ola Mahmoud, Shailee Pradhan, we well as the anonymous referees and the JFDS editor for their comments and suggestions. We extend our acknowledgements to different participants of the risk seminar at UC Berkeley, UCSB, and other seminars at the Universities of Zurich and Basel. We are grateful to the Swiss Re Institute for funding through the Consortium for Data Analytics in Risk (CDAR), and to OWL Analytics and TruValue Labs for data.}  \endgraf\bigskip\bigskip
  Published at the {\textit{Journal of Finance and Data Science}} \endgraf}}
}
\begin{document}
\maketitle 
\begin{abstract}
\noindent We use supervised learning to identify factors that predict the cross-section of returns and maximum drawdown for stocks in the US equity market. Our data run from January 1970 to December 2019 and our analysis includes ordinary least squares, penalized linear regressions, tree-based models, and neural networks. We find that the most important predictors tended to be consistent across models, and that non-linear models had better predictive power than linear models. Predictive power was higher in calm periods than in stressed periods.  Environmental, social, and governance indicators marginally impacted the predictive power of non-linear models in our data, despite their negative correlation with maximum drawdown and positive correlation with returns. Upon exploring whether ESG variables are captured by some models, we find that ESG data contribute to the prediction nonetheless.
\end{abstract}

\newpage
\section{Introduction}
We apply a variety of supervised learning models to forecast returns and maximum drawdown---the largest decline in cumulative return over a fixed period. As predictors, we use standard accounting ratios and factors, sector information, and environmental, social, and governance (ESG) indicators. We look specifically at whether or not ESG indicators augment the predictive power of our methods.
\subsection{Motivation}
The search for factors that drive the return and risk of individual stocks has been the focus of the asset pricing theory for decades. While the capital asset pricing model (CAPM), see \cite{Treynor1961}, \cite{Sharpe1964}, \cite{Lintner1965} and \cite{Mossin1966}, is probably the first model to address this question and states that stocks' returns are explained by their market risk through the betas, empirical findings suggested that this model is incomplete. Despite decades of research and hundreds of papers, a thorough understanding of the drivers of cross-sectional variation of expected returns eludes us. The Fama and French three-factor model, see \cite{FamaF1992}, and the Carhart four-factor model, see \cite{Carhart1997}, are well-accepted academic models used to capture this variation. They demonstrate that size, book-to-market ratio, and momentum are significant drivers of asset returns, and they complement the CAPM. Nevertheless, new empirical studies continue to emerge, and document a growing number of factors, see \cite{FamaF2008} and \cite{HarveyLZ2015} for an overview, or \cite{GreenHZ2013} and \cite{FengGX2017} for extensive factors mining. In recent papers, \cite{GuKX2018, chen2020deep} investigate which firm characteristics predict the cross-section of expected return using machine learning and deep learning techniques, and find an advantage in using non-linear models.\\
\\
Firm characteristics are also used to forecast risk. For example, Barra models, which are standard in industry, use firm characteristics to estimate stock covariance matrices, see \cite{barra2014}. They decompose the covariance matrix of security returns using the estimation of sensitivities of the securities to the common factors, the covariance matrix of factors, and the variances of security-specific returns. \cite{Vozlyublennaia2013} investigate the effect of firm characteristics in the dynamics of idiosyncratic risk. They find that firm characteristics can be used in the analysis of the differences in risk across securities. \cite{HerkosvicKL2013} show that a firm's idiosyncratic volatility obeys a strong factor structure. 
In addition to returns, our paper is interested in an under-looked but yet very important metric to assess risk by portfolio managers; maximum drawdown.\\
\\
Maximum drawdown has also received extensive attention in the literature on financial markets. However, most of the papers are theoretical. 
\cite{Taylor75} elucidates properties of maximum drawdown under the assumption of a Brownian diffusion for the underlying process, \cite{MagdonIsmailAAP2004} on the other hand give a series representation of the maximum drawdown distribution and estimate its expected value. A second body of literature addresses portfolio construction  and formalizes some risk measures based on maximum drawdown, see \cite{ChekhlovUZ2004}, \cite{HeidornKR2009}, or \cite{GoldbergM2016}. \\
\\
In many cases, investors are forced to liquidate positions after a large loss. A body of literature uses ranking rules to momentum-style portfolio construction. In this concept, tail risk is considered in risk-adjusted performance measure like the Calmar ratio. \cite{rachev2007} report that tail behavior can predict future price directions of equities. They report that by using risk-adjusted performance measure, namely the Stable-Tail Adjusted Return ratio. \cite{jaehyung2021} use maximum drawdown and its consecutive recovery as a stocks selection rule and find that it outperforms other alternative momentum portfolios. Such strategies resulted in a better risk-adjusted returns. Close to this line of inquiry is \cite{DanielM2016}, who look at how momentum strategies perform following drawdown, and \cite{Plastira14}, who studies performance rankings of popular size, value, reversal and momentum portfolios using drawdown-based portfolio performance measures.\\
\\
This project began with the question of whether sustainability metrics predict stock performance. Sustainability metrics include a set of scores that specialized agencies assign to companies based on their environmental, social, and governance (ESG) strengths and weaknesses.  Although sustainable investing was a niche topic a few years ago, it has become mainstream for many asset managers and mutual funds.  
According to the US SIF Foundation’s 2020 Report on US Sustainable and Impact Investing Trends, as of year-end 2019, one out of every three dollars under professional management in the United States—\$17.1 trillion—was managed according to sustainable investing strategies. 
It is no surprise that it has also become a key topic in financial research. SRI can be roughly defined as ``an investment process that integrates social, environmental, and ethical considerations into investment decision making'', see \cite{Renneboog2008}. This involves screening companies for corporate social responsibility on the basis of environmental (E), social (S), and governance (G) criteria.  Whether or not ESG and other non-financial criteria have a material economic impact is the subject of an ongoing debate. \\
\\
The main question that animates the dialog around ethical investing is whether companies that ``do good" also ``do well." This statement suggests that companies with superior ESG performance generate higher financial performance and/or have a lower risk. In that sense, a wide range of literature explores the link between sustainability metrics and several dimensions of performance and risk. \cite{PrestonO1997} explore the possible empirical association between social and financial performance (through return on assets and other measures) in longitudinal data. Their sustainability data are based on a \textit{Fortune} magazine reputation ratings for individual companies. They find that positive synergies explain social-financial performance correlations. \cite{KhanSY2016} explore the question of materiality, i.e., a classification that maps different sustainable indicators as material for different industries. Their analysis relies on KLD\footnote{Founded in 1988, KLD Research \& Analytics, Inc. provides performance benchmarks, corporate accountability research, and consulting services. The company offers environmental, social, and governance research for institutional investors.} data and the ``materiality'' mapping provided by the Sustainability Accounting Standards Board (SASB)\footnote{The Sustainability Accounting Standards Board (SASB) was founded in 2011 to develop and disseminate sustainability accounting standards.}.
Using both portfolio and firm-level regressions, they find that firms with good ratings significantly outperform firms with poor ratings. \textit{Immaterial} sustainability issues, on the other hand, do not improve financial performance. Using MSCI ESG KLD STATS data between 2000 and 2016 on the US market, \cite{brogi2019environmental} also find evidence of the positive impact of ESG on the return of assets (ROA), especially in the banking sector. \cite{giese2019foundations} find a positive impact on companies' valuation and performance through reduced capital costs, higher valuations, higher profitability, and lower exposure to tail risk. The analysis of over 2200 individual stocks by \cite{friede2015esg} highlights that about 90\% of them show a non-negative relationship between ESG and corporate financial performance. \cite{halbritter2015wages} show that the direction of the overperformance of ESG portfolios is strongly dependent on the rating providers.\\
\\
An opposing point of view is ``doing good but not well". This view is linked to the ``managerial opportunism hypothesis," which suggests that managers tend to maximize their gains and that socially responsible activity may cost resources to the firm, putting it at a disadvantage relative to firms that invest less in sustainability issues, see \cite{AupperleCH1985}. In that sense, \cite{BrammerBP2006} and \cite{HongK2009} show a higher performance for portfolios with low sustainability performance compared to their peers with high sustainability performance. In a recent paper, \cite{bruno2022honey} finds that most of the outperformance of ESG-based strategies can be explained by their exposure to equity-style factors that are mechanically constructed from balance sheet information. 
\cite{madhavan2021} separate the impact of ESG variables into a factor-related effect and an idiosyncratic effect. They found that the factor-related effect drives the alpha and active returns, but failed to
reject the lack of relationship between idiosyncratic ESG components and performance. \cite{madhavan2020factor} find similar results for fixed-income mutual funds. They conclude that ESG funds derive a significant portion of their performance from traditional factors, such as quality and low volatility.
Other studies on the European market find similar results, see \cite{su12166387}. \cite{billio2021inside} analyze the ESG rating criteria used by known agencies. They find that the disagreement in the scores among agencies disperses the effect of the preferences of ESG investors and dissipates their impact on financial performances. As a result, a new body of literature studies the rating disagreement and incorporates it in the performance analysis, see \cite{gibson2021esg, avramov2022sustainable}.\\
\\
Our paper, on the other hand, explores the question of how sustainability issues impact financial risk through maximum drawdown. We ask whether ESG data enhance our ability to predict future performance and risk. Previous literature explores similar questions by testing for a relationship between sustainability measures and firm systematic risk. In this line of inquiry,  \cite{JoN2012} and \cite{BenlemlihSQG2018} use the CAPM beta as the measure of a firm's systematic risk. While the first of the two papers focuses on `controversial industries', both studies find a strong negative link between corporate social performance and systematic risk. \cite{Chollet2018} consider, in addition to systematic risk, specific and total risk, translated respectively by the standard deviation of residuals from the CAPM model and the volatility of stocks. Using Thomson Reuters ASSET4 to measure ESG scores, they find that a firm's good social and governance performance reduces its financial risk. 

\subsection{Contributions}
While this research began with the assessment of  ESG's impact on stock performance, the contributions of this paper go beyond that. We identify determinants of the cross-section of middle-term returns and drawdown using firms' quantitative and qualitative characteristics. We perform a pooled regression analysis using firm-level factors and nine regression algorithms in supervised statistical learning and compare the performance of these methods based on their out-of-sample performance.\\
\\
Stock returns are difficult to predict in general and particularly in the short term and incorporate a low signal-to-noise ratio but long-term investors rely on fundamentals for their stock selection. Maximum drawdown, which measures a firm's risk of successive negative returns over a period of time, has higher predictability across firms. The first analysis seeks to find if stock level characteristics explain long-term returns and maximum drawdown over a large data set that includes 2008 financials without ESG variables. We then split our data to account for ESG variables and repeat the analysis. We find that maximum drawdown gives high predictability results in both cases compared to returns. In both cases, we can obtain both high predictability and economically meaningful results.\\
\\
Our empirical analysis clarifies what features are most valuable to forecasting a company's future return and drawdown. By applying nine regression methods, we can identify a consistent set of variables that effectively forecast maximum drawdown and, to a certain extent, returns.
\subsection{Empirical findings}
{We rely on monthly stock data from January 1970 until December 2019 with a total of 600 months. After some data processing and focusing on companies listed on Russel 3,000, we obtain an average of 2,717 individual stocks per month. Our predictions are based on 119 lagged firm characteristics and an additional 7 ESG-related variables. Our first analysis uses only non-ESG variables for a long test period.  Our second analysis is constrained by the ESG data history, so we adjust the training and test sets accordingly. Due to the co-linearity between OWL Analytics ESG scores, we group them into three sets; one where we include granular indicators,  a second where we include E, S, and G, and a third with the single, fully aggregated ESG score but provide results for the E, S, G, and ESG scores only. TruValue Labs's ESG indicators are also added, both separately and in combination with OWL data. Finally, we compare the forecasts of a set of linear and non-linear supervised learning algorithms. The empirical findings of our analysis are summarized in the following.}
\noindent \textit{Machine learning offers new tools to address regression problems and account for non-linear relationships in asset pricing theory.} Supervised learning approaches usually aim to predict an output from several input variables. Linear regressions, either ordinary or penalized, and dimension reduction methods like principal components and partial least square regressions are limited to linear relationships. The advantage of more complex models used in machine learning like tree-based models and neural networks is that they can overcome this limitation and account for non-linear relationships without increasing the dimension of the problem (by introducing new variables). Our results show that non-linear models improve performance. However, linear models are also a match when data are primarily processed, generally offer comparable performance, and are robust when little data are available. \\
\\
\textit{The ranking of one-year returns and maximum drawdown across stocks was predictable in our out-of-sample test period. Maximum drawdown has a better prediction performance than the returns.}  We report as a performance metric the mean squared error (MSE) and Kendall rank correlation, and we analyze decile portfolios based on predicted returns and maximum drawdown. While MSE assesses how small the error of the prediction to the true value is, the Kendall correlation, which measures the concordance between the prediction and the realization, gives a better results interpretation. In particular, a large positive value would suggest that your strategy is not ranking stocks randomly. Kendall correlation on the test period is close to 50\% for maximum drawdown and over 10\% for log excess returns (logER).
The cross-sectional correlation reached its lowest value during the 2008 crisis, $\sim 20\%$ for logER and, $\sim 25\%$ for the maximum drawdown. During mild market conditions, the value exceeded over 20\% for logER and  $55\%$ for MDD. The positive sign of the Kendall correlation suggests that even during low-performance periods, there was a positive relationship between the realized and predicted values of $MDD$ while return prediction can lead to bad investment choices.\\
\\
\textit{The set of dominant features was largely consistent across models.} Based on a sensitivity analysis that ranks variables by their predictive power, when excluding ESG, the top five variables in most models were volatility-based measures (one-year, one-month, and idiosyncratic), bid-ask spread, and size. Among important variables, we also find momentum and earnings-to-price.\\
\\
\textit{Some combinations of environmental, social, and governance scores improved the predictability of returns and maximum drawdown, although marginally, when added to the top-performing non-ESG variables over the test period, January 2015 to December 2019.} We find that penalized linear regression models discarded these variables but that non-linear models captured them. In fact, through exploring variable importance using the Shapley value, we find that the G score and S are among the top features in the prediction and have a favorable relationship to risk (negative correlation with MDD) and performance (positive correlation with returns). The results we find suggest that the correlation fades away when controlling for firm characteristics. 
\color{black}
\newpage
\section{Methodology}\label{sec:methodology}
One of the goals of this paper is to compare the ability of different supervised learning methods to predict next-period return and maximum drawdown, or more precisely, their ranking across stocks, from an observed set of quantitative and qualitative firm attributes. We summarize the (well-known) methods used in the analysis here for completeness.\\
\\
All the methods we consider follow the general equations:
\begin{align}\label{eq:themodel0}  
y_{i, t+1} &= \mathbb{E}_t[y_{i, t+1}] +  \varepsilon_{i, t+1},
\end{align}
and 
\begin{align}\label{eq:themodel1}  
\mathbb{E}_t[y_{i, t+1}] &=  f^*(x_{i, t}),
\end{align}
where $f^*$ defines a general mapping function, $y_{i t+1}$ denotes the dependent variable, i.e., log excess return or maximum drawdown, of firm $i$ over the period from time $t$ to time $t+1$, $x_{i, t} = [x_{i, t}^1, ..., x_{i, t}^J]$ is a vector of realizations of firm characteristics for a stock $i$ at time $t$, and $\varepsilon_{i, t+1}$ an error term.\\ 
\\
The relationship expressed in formulas (\ref{eq:themodel0}) and (\ref{eq:themodel1}) mirrors the asset pricing relationships specified in countless papers in the finance literature.  In those papers, the dependent variable is usually excess return over the risk-free rate or another benchmark.
\subsection{The Models}
\subsubsection*{Linear models}
The first set of methods we use are linear regression models, in which the mapping function $f$ is a linear combination of entries $\boldsymbol{x_{i, t}} = [x_{i, t}^1, x_{i, t}^2, ..., x_{i, t}^J]$:
\begin{align*}
f(x_{i, t}; \beta) &= \boldsymbol{x_{i, t}'}\boldsymbol{\beta} \\
& =  \beta_0 + \sum_{j=1}^{J}\beta_j x_{i, t}^{j}
\end{align*}
where the $\beta$ is the parameter vector and $ \boldsymbol{x_{i, t}'}$ is the transpose of $ \boldsymbol{x_{i, t}}$ the vector of the firm characteristics for stock $i$ at time $t$.

\subsubsection*{Ordinary Least Square Regression}
To find the regression parameters $\beta_0, \beta_1, ..., \beta_J$, linear regression minimizes the sum of squares errors between the observation $y_{i, t+1}$, and the prediction $f(x_{i, t})$:
\begin{align*}
\beta = \argmin_{\beta} \frac{1}{NT}\sum_{t=1}^{T}\sum_{i=1}^{N_t}\big(y_{i, t+1} - f(x_{i, t+1})\big)^2
\end{align*}
When the number of input features is large or when the regressors are correlated, linear regression can lead to overfitting and leads to spurious coefficients. 

\subsubsection*{Penalized regression techniques: Lasso, Ridge, and Elastic Net}
Penalized regression models aim to create a more robust output model in the presence of a large number of potentially correlated variables. They are a simple modification of the ordinary linear regression that introduces a regularization term in the optimization problem - a technique in Machine Learning that aims to reduce the out-of-sample forecasting error by penalizing the coefficients. In this paper, we explore the three main ones: Lasso, Ridge, and Elastic Net.
\begin{itemize}
\item The first penalized regression technique, called \textit{Lasso}, for ``least absolute shrinkage and selection operator'', see \cite{Tibshirani1996}, is based on a penalty term equal to the absolute value of the beta coefficients:
		\begin{align*}
			\beta = \argmin_{\beta}\frac{1}{NT}\sum_{t = 1}^{T}\sum_{i = 1}^{N_t}\big(y_{i, t+1} - f(\boldsymbol{x_{i, t}}; \beta)\big)^2 + \lambda \sum_{j = 1}^{J}\mid\beta_j\mid
		\end{align*}
		where $\lambda$ is a non-negative hyperparameter.
\item Ridge regression, see \cite{Hoerl1962}, adds a penalty related to the square of the magnitude of the coefficients called $\ell_2$ regularization and solves the following objective function:
		\begin{align*}
			\beta = \argmin_{\beta}\frac{1}{NT}\sum_{t = 1}^{T}\sum_{i = 1}^{N_t}\big(y_{i, t+1} - f(\boldsymbol{x_{i, t}}; \beta)\big)^2 + \lambda \sum_{j = 1}^{J}\beta^2_j
		\end{align*}
	where $\lambda$ is a non-negative hyperparameter.
\item Elastic Net, see \cite{ZouH2005}, uses an intermediate objective function between the Lasso and Ridge:
		\begin{align*}
			\beta = \argmin_{\beta}\frac{1}{NT}\sum_{t = 1}^{T}\sum_{i = 1}^{N_t}\big(y_{i, t+1} - f(\boldsymbol{x_{i, t}}; \beta)\big)^2 + \lambda_1 \sum_{j = 1}^{J}\mid\beta_j\mid + \lambda_1\sum_{j = 1}^{J}\beta^2_j
		\end{align*}
		where $\lambda_1$, $\lambda_2$  are two non-negative hyperparameters.
\end{itemize}
The ordinary least squares objective is obtained by setting the parameter $\lambda=0$ (resp. $\lambda_1 = \lambda_2=0$). Moreover, as $\lambda$ (resp. $\lambda_1$ and $\lambda_2$) increases, we choose a smaller set of predictors by decreasing the value of the coefficients and shrinking the least relevant ones.

\subsubsection*{Dimension Reduction}
Dimension reduction techniques aim to decrease the number of features in a data set without discarding salient information. In contrast to penalized regression methods, which discard weak regressors by setting their loadings to zero, dimension reduction techniques form an uncorrelated linear combination of the predictors for the purpose of reducing noise and concentrating signal. In our analysis, we rely on two widely used methods, principal component regression, and partial least squares.\\
\\
Principal Component Regression (PCR) is a two-step procedure.  The first step is a principal component analysis (PCA) that combines the independent variables into a set of leading components ranked by their explained variance. The predicted variable plays no role in this step.  The second step is a simple linear regression on the leading components. PCA is one of the most widely used dimensionality reduction techniques, and it dates back to \cite{Pearson1901}. The key idea is to find a new coordinate system in which the input data can be expressed with fewer variables without a significant error.\\
\\
Unlike PCR, in which the two steps are performed separately, a partial least squares (PLS) regression combines dimension reduction and regression by directly taking into account the covariance of the predictors with the target prediction. This is carried out by estimating, for each predictor $p$, a univariate return prediction coefficient via OLS. This coefficient $\varphi_p$ represents the partial sensitivity of returns to a given predictor $p$. Predictors are then averaged into a single aggregate component with weights proportional to $\varphi_p$, which would give the strongest univariate predictors the highest weights. Then the target and all predictors are orthogonalized with respect to previously constructed components, and the procedure is repeated on the orthogonalized set. The procedure stops when the desired number of components is obtained.\\
\\
More formally, we write the linear model in its vectorized version,
\begin{align*}
R = Z\theta + E,
\end{align*}
where $R$ is the $NT\times 1$ vector of $r_{i, t+1}$, $Z$ is the $NT\times P$ matrix of stacked predictors $z_{i, t}$, and $E$ is a $NT\times 1$ vector of residuals $\varepsilon_{i, t+1}$.\\
\\
The linear model given above is re-written for the set of reduced predictors:
\begin{align*}
R = (Z\Omega_K)\theta_K + E,
\end{align*}
where $K$ is the number of reduced predictors corresponding to a linear combination of the initial ones, $\Omega_K$ is the $P\times K$ matrix with columns $w_1, w_2, ..., w_K$, where $w_j$ for $j \in 1,..., K$ is the set of linear combination weights used to create the $j^{th}$ predictive components, and $Z\Omega_K$ is the reduced version of the original predictor set.

\subsubsection*{Regression Trees and Random Forests}
Decision trees are among the simplest non-linear models that rely on a tree structure to approach the outcome. We can express the prediction of a tree with $M$ terminal nodes and depth $L$ as:
\begin{align*}
f(\boldsymbol{x_{i, t}}; \theta, M, L) = \sum_{m=1}^{M}{\theta_m}\mathds{1}_{\boldsymbol{x_{i, t}} \in C_m(L)},
\end{align*}
where each $C_m(L)$ is one of the $M$ partitions of the data.\\
\\
The algorithm that fits the decision sequentially splits subsets of the data in two on the basis of one of the predictors.  The split at each step is chosen to optimize a loss function defined in terms of impurities in the child nodes. Impurity is typically measured in terms of the Gini index or entropy. To prevent overfitting and to ensure the tree is interpretable, different criteria can be used like the maximum depth of the tree or node size.\\
\\
A random forest averages the output of many decision trees. Each decision tree is fit on a small subset of training examples or constrained to use a subset of input features. Doing so increases the bias relative to a simple decision tree, but decreases the variance; see \cite{Breiman2001} for more details.\\
\\
Formally, if a regression tree has $M$ leaves and accepts a vector of size $n$ as input, then we can define a function $q: \mathbb{R}^n\rightarrow\{1, ..., T\}$ that maps an input to a leaf index. If we denote the score at a leaf by the function $w$, then we can define the $k$-th tree (within the ensemble of trees considered) as a function $f_k(x) = w_{q(x)}$, where $w \in \mathbb{R}^T$. For a training set of size $n$ with samples given by $(x_i, y_i)$, $x_i \in \mathbb{R}^m$, $y_i \in \mathbb{R}$, a tree ensemble model uses $K$ additive functions to predict the output as follows:
 \begin{align*}
\hat{y}_{i, t+1} = f(\boldsymbol{x_{i, t}}) = \sum_{k=1}^{K}{f_k(\boldsymbol{x_{i, t}})}.
\end{align*}
\subsubsection*{Extreme Gradient Boosting}
The term `boosting' refers to the technique of iteratively combining weak learners (i.e., algorithms with weak predictive power) to form an algorithm with strong predictive power. Boosting starts with a weak learner like a regression tree algorithm, and records the error between the learner's predictions and the actual output. At each stage of the iteration, it uses the error to improve the weak learner from the previous iteration step. If the error term is calculated as a negative gradient of a loss function, the method is called `gradient boosting.' Extreme gradient boosting (or XGBoost) refers to the optimized implementation in \cite{ChenG16}. \\
\\
Formally, the model uses $K$ additive functions to predict the output as follows:
\begin{align*}
\hat{y}_{i, t+1} = f(\boldsymbol{x_{i, t}}) = \sum_{k=1}^{K}{f_k(\boldsymbol{x_{i, t}})},
\end{align*}
where we take $f_k(x) = w_{q(x)}$ ($q : \mathbb{R}^m \rightarrow T, w \in \mathbb{R}$) from the space of regression tree. The function $q$ represents the structure of each tree that maps an example of the data set to the corresponding leaf index, $T$ is the number of leaves in the tree, and each $f_k$ corresponds to an independent tree structure $q$ and leaf weight $w$.\\
\\
To learn the set of functions in the model, the regularized objective is defined as:
\begin{align*}
\mathcal{L}(f) = \sum_{t}\sum_{i}l(\hat{y}_{i, t+1}, {y}_{i, t+1}) + \sum_{k}\Omega(f_k)
\end{align*}
where $\Omega(f) = \gamma T + \frac{1}{2}\lambda\norm{w}^2$.\\
\\
The model is then optimized in an additive manner. If $\hat{y}_i^{(t)}$ is the prediction for the $i$-th training example at the $t$-th stage of boosting iteration, then we seek to augment our ensemble collection of trees by a function $f_t$ that minimizes the following objective:
\begin{align*}
\mathcal{L}^{(t)} = \sum_{i=1}^{n}l\big(y_i, \hat{y}_i^{(t-1)} + f_t(x_i)\big) + \Omega(f_t).
\end{align*}
The objective function is approximated by a second-order Taylor expansion and then optimized (see \cite{ChenG16} for details and calculation steps). To prevent overfitting, XGBoost uses shrinkage and feature sub-sampling.
\subsubsection*{Artificial Neural Networks: The Multi-Layer Perceptron}
Artificial neural networks (ANN) are a set of algorithms inspired by the human brain and designed to recognize patterns. The idea behind such a framework is to represent complex non-linear functions by connecting simple processing units into a \textit{neural network}, each of which computes a linear function, possibly followed by a non-linearity.\\
\\
A neuron-like processing unit is given by:
\begin{align*}
a = \phi\big(\sum_j w_j x_j + b \big),
\end{align*}
where the $x_j$'s are the inputs to the unit, the $w_j$'s are the weights, $b$ is the bias, $\phi$ is a non-linear activation function, and $a$ is the unit's activation. An activation function is a function used to transform a trigger level of a unit (neuron) into an output signal. Examples of such activation functions are:
\begin{itemize}
	\item The identity activation function: $\phi(x) = x$.
	\item The logistic activation function:  $\phi(x) = \frac{1}{(1 + e^{-x})}$.
	\item The hyperbolic tan function `Tanh': $f(x) = \tanh(x)$.
	\item The rectifier linear unit function `ReLu': $f(x) = \max(0, x)$.
\end{itemize}
The combination of a set of these units is what forms the neural network. Each unit performs a simple function, but in aggregate, the units can do more complex computations. In our analysis we apply a variant of the simple \textit{feed-forward neural network}, that is, the multi-layer perceptron. In such a model, the units are arranged in an acyclic graph, and calculations are done sequentially (in contrast to a \textit{recurrent neural network}, where the graph can have cycles).\\
\\
The multi-layer perceptron (MLP), as shown in Figure \ref{fig:multilayerperceptron}, is composed of a set of layers, each of which contains identical units. In an MLP, the network is fully connected, i.e., every unit in one layer is connected to every unit in the next layer. The {\it input} layer takes the values of the predictors. The last {\it output} layer has a single unit in the case of regression. The intervening {\it hidden} layers are mysterious because we do not know ahead of time what their units should compute. The number of layers is known as the depth, and the number of units in a layer is known as the width. ``Deep learning" refers to training neural networks with many layers.\\
\\
\begin{figure}
\centering
\def\layersep{2.5cm}
\begin{tikzpicture}[shorten >=1pt,->,draw=black!50, node distance=\layersep]
    \tikzstyle{every pin edge}=[<-,shorten <=1pt]
    \tikzstyle{neuron}=[circle,fill=black!25,minimum size=17pt,inner sep=0pt]
    \tikzstyle{input neuron}=[neuron, fill=green!50];
    \tikzstyle{output neuron}=[neuron, fill=red!50];
    \tikzstyle{hidden neuron}=[neuron, fill=blue!50];
    \tikzstyle{annot} = [text width=4em, text centered]

    \foreach \name / \y in {1,...,4}
        \node[input neuron, pin=left:$x_\y$] (I-\name) at (0,-\y) {};

    \foreach \name / \y in {1,...,5}
        \path[yshift=0.5cm]
            node[hidden neuron] (H-\name) at (\layersep,-\y cm) {};

    \node[output neuron,pin={[pin edge={->}]right:$y$}, right of=H-3] (O) {};

    \foreach \source in {1,...,4}
        \foreach \dest in {1,...,5}
            \path (I-\source) edge (H-\dest);

    \foreach \source in {1,...,5}
        \path (H-\source) edge (O);

    \node[annot,above of=H-1, node distance=1cm] (hl) {Hidden layer};
    \node[annot,left of=hl] {Input layer};
    \node[annot,right of=hl] {Output layer};
\end{tikzpicture}
\caption{An Artificial Neural Network with one hidden layer}
\label{fig:multilayerperceptron}
\end{figure}

\noindent We denote the input units by $x_j$, the output unit by $y$, and  the $\ell$-th hidden layer by $h_i^{(\ell)}$. Since MLP is fully connected, each unit receives input from all the units in the previous layer. This implies that each unit has its own bias and a weight is associated with every pair of units in consecutive layers:
\begin{align*}
h_i^{(1)} &= \phi^{(1)}\Big(\sum_j w_{ij}^{(1)}x_j + b^{(1)}_i \Big) \\
h_i^{(1)} &= \phi^{(2)}\Big(\sum_j w_{ij}^{(2)}h_j^{(1)} + b^{(2)}_i \Big) \\
y_i &= \phi^{(3)}\Big( \sum_j w_{ij}^{(3)}h_j^{(2)}+ b^{(3)}_i \Big),
\end{align*}
where $\phi^{(1)}$ and $\phi^{(2)}$ are the activation functions (which can be different for different layers).
\newpage
\subsection{The Data}
\subsubsection*{The independent variables}
\begin{figure}[!h]
\caption{Number of stocks in time between January 1970 and December 2019.}
\centering
\includegraphics[scale=0.5]{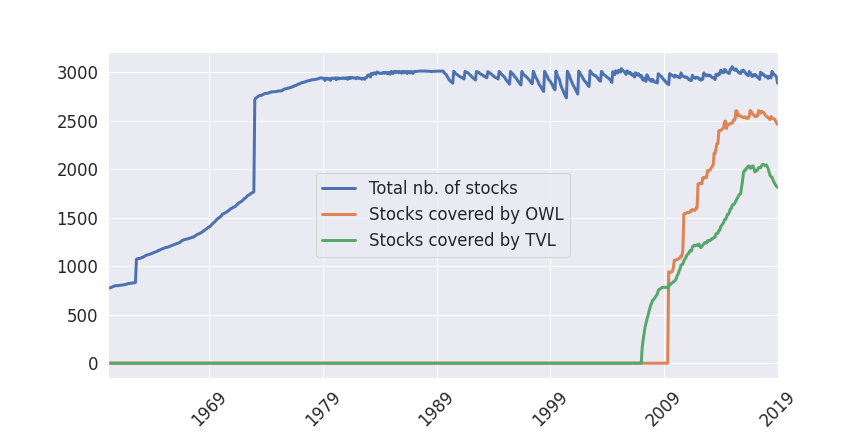}
\label{fig:NB_Stocks}
\end{figure}

The independent variables in the analysis include firm characteristics taken from the CRSP/Compustat database. The construction of firm characteristics and the notation we use to refer to them are taken from Wharton Research Data Service (WRDS) with some further cleaning.
Descriptions of the firm characteristics are provided in the appendix. We work with CRSP stocks, identified by their PERMNO code. We restrict the analysis to shares within the Russel 3000 to limit the impact of micro and small caps. CRSP firm characteristics are then merged with ESG data from two different data providers, OWL Analytics and TruValue Labs.\\
\\
OWL Analytics aggregate data from different providers to generate their ESG scores, which are updated on a monthly basis and cover 12 primary categories:
\begin{itemize}
	\item 3 environment scores: pollution prevention (E1), environmental transparency (E2), resource efficiency (E3).
	\item 6 social scores: compensation \& satisfaction (EMP1), diversity \& rights (EMP2), education \& work condition (EMP3), community \& charity (CIT1), human rights(CIT2) and sustainability integration (CIT3).
	\item 3 governance scores: board effectiveness (GOV1), management ethics (GOV2), and disclosure \& accountability (GOV3).
\end{itemize}
\noindent The scores of each category are aggregated into the main three scores E (for environment), S (for social), and G (for governance), before being averaged into a single ESG metric (ESG score). Stocks in the OWL Analytics database are identified by their ISIN, and their coverage starts in 2009-07. We limit the current analysis to the pillars E, S, and G and the aggregated ESG score from Owl Analytics. \\
\\
TruValue Labs (TVL), on the other hand, relies on public data, mainly news, and filters the data through the SASB ``materiality map'' to assess their ESG impact. Their data coverage starts in 2007. In our analysis, we focus on four of their scores: 
    \begin{itemize}
        \item Trailing 12-month volume: Number of articles tagged to SASB categories during the past 12 months.
        \item Insight: Exponentially-weighted moving average (EWMA) of pulse score as a long-term  sentiment indicator.
        \item Momentum: Slope of Insight score over past twelve months intended to identify companies with improving or deteriorating ESG.
    \end{itemize}
ESG data and firm characteristics are merged to form our data set. WRDS data are usually available by the end of the month and made public with a lag. OWL Analytics ESG scores require a two months lag, while TruValue Labs are lagged by one day since they are updated daily. Those sets of predictors are then aligned with the one-year forward maximum drawdown. \\
\\
We exclude stocks and dates with a lot of missing data and winsorize variables beyond the median by 3 interquartiles to the 1\% and 99\% percentile value cross-sectionally. 
Finally, we keep only stocks for which returns are available for the subsequent twelve months (necessary to calculate one-year logER and MDD without any missing dates). The resulting  data set covers 564 months (from 1970-01 until 2019-12) with a total of $14,488$ stocks and an average of $2722$ stocks per date. Figure \ref{fig:NB_Stocks} shows the number of stocks available in the analysis for each time period, along with the number of stocks with OWL Analytics and TruValue Labs data.
\color{black}
\subsubsection*{The dependent variable}
\begin{figure}[h]
\caption{One-year forward Maximum Drawdown}
\caption*{The two figures report the evolution in time of the one-year forward MDD for IBM and Morgan Stanley between 1990-01 and 2019-12 (left), and the cross-sectional distribution of MDD for all stocks on 2008-01-01 and 2014-01-01.}
\centering
\includegraphics[width=8cm, height=6cm]{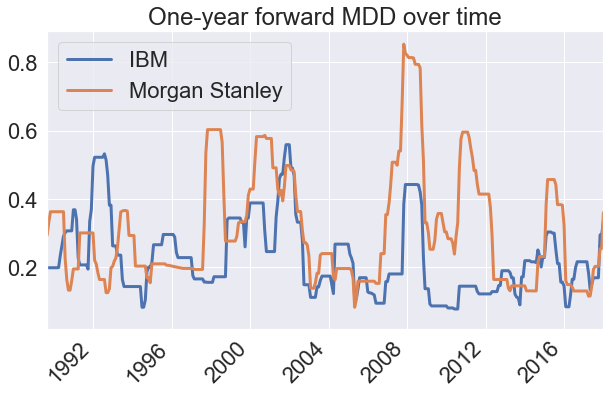}
\includegraphics[width=8cm, height=6cm]{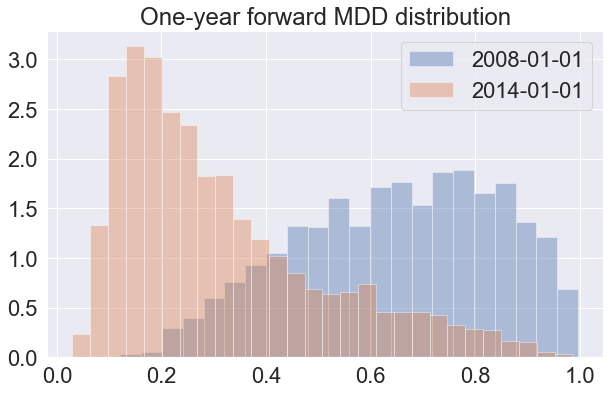}
\label{fig:MDD_time_distribution}
\end{figure}

As reported to Benjamin Graham\footnote{Warren Buffet claims his mentor Benjamin Graham wrote it, however, the quote is not found in either of Graham's books, Security Analysis or The Intelligent Investor.}; ``in the short run, the market is a voting machine, but in the long run, it is a weighing machine''. Motivated by the importance of long-horizon predictions of stocks risk and return, and probably a tendency to be predictable by stocks' fundamental data, we orient the exercise of factor models using statistical regression methods to a one-year horizon.\\
\\
Because of the desirable properties of log returns, we use log excess returns to the risk-free rate. Also, rather than using the volatility or variance as the risk measure, we focus on an under-looked, but nonetheless important risk measure, which is maximum drawdown.\\
\\
The maximum drawdown of a stock is defined as its largest cumulative loss from peak to trough over a fixed period $\tau$. Letting $P$ denote stock price, maximum drawdown can be calculated through the following formula\footnote{While maximum drawdown is a negative return, we take the form that gives a positive value/loss.}:
\begin{align*}
MDD(P) = \sup_{t\in [0, \tau]}\sup_{s\in [t, \tau]}\Big(\frac{P_t - P_s}{P_t}\Big).
\end{align*}
By construction, the maximum drawdown is unique for each period (and each stock), however, it depends on the observation frequency. For example, using daily observations, the calculated maximum drawdown would miss the flash crash that happened in 2010. Intraday observations on the minutes or seconds level in this case would detect a larger maximum drawdown than what daily observations do. However, for several reasons such as the availability of intraday data, and the unlikelihood that such uncommon market activity is driven by fundamentals, we stick to daily observations for maximum drawdown calculations in this paper.  We rely on a one-year forward maximum drawdown using the same moving window we use to calculate log excess returns. We display in Figure \ref{fig:MDD_time_distribution} two representations of MDD: its evolution in time for two stocks and its cross-sectional distribution for two periods.
\newpage
\subsubsection*{Training, validation, and test data}
 In financial economics, it is standard practice to split a data set into in-sample and out-of-sample subsets and to evaluate  performance based on the latter. For machine learning models, it is more common to split the data into three subsets:  training, validation, and test.  The training set is used to fit a number of models with the same general architecture but different hyperparameters.\footnote{In machine learning, a hyperparameter is a  parameter whose value is set before the learning process begins, and is specific to the model's architecture.  Examples of hyperparameters are the penalization term for a penalized regression and the number of hidden layers and number of units in a neural network. Hyperparameters are specified through a validation process and are chosen to minimize model error over the validation set.} The validation set is used to tune the hyperparameters.  The test set is used to evaluate model performance.  This more elaborate procedure addresses the complexity of machine learning models, and their  tendency to overfit or underfit when hyperparameters are poorly chosen\footnote{$K$-fold cross-validation is a standard method for tuning hyperparameters.  The method splits training data into $k$ equally sized subsets.  A model with a fixed set of hyperparameters is fit on the union of $k-1$ subsets and evaluated on the $k$th subset. This process is repeated $k-1$ times and the evaluation scores are averaged, yielding an overall score. This practice is, however, very costly and does not respect the time ordering of events, which is essential to our analysis.}.\\
\\
The avoidance of look-ahead bias, the goal of subjecting our models to the most challenging tests  possible, and the availability of data guide the specification of our training, validation, and test sets. 
To select the parameters of the models, we omit ESG variables and split the data into a training set that runs from 1970-01 until 1990-12 and a validation set that runs from 1991-12 until 2001-01. Once the best model parameters selected, we then retrain the model and test it. First we do so without incorporating ESG variables due to their short history. In this specific case, our training data runs between 1970-01 and 2000-01-01.\\
\\
The short history of ESG data (which begin on 2009-07 for OWL and 2007-01 for TVL) forces us to make two compromises). First, to limit the effect of missing data we dismiss data points without any ESG information. We also opt for a test set of one year at a time and increase the training set (fixed starting data and moving ending date with retraining at each beginning of the year). Therefore, we train the models 5 times with a starting date on 2009-01 and an end date from 2014-01 to 2018-01 with one year increment. Each trained model will be used to predict one year length test set between 2015 and 2019\footnote{Since we evaluate maximum drawdown over one-year periods and stagger start dates by a month, a period of $N$ years contains $12\times (N-1)$ observations.}.

\subsubsection*{Preprocessing}
Data preprocessing is the technique of transforming raw data into a machine-understandable format. Some machine learning algorithms are particularly sensitive to ranges of input values and stationarity of the variables, and rescaling is a common way to overcome this issue. However, there is no consensus on a rescaling method to apply to these problems although it is very common to see z-scoring and uniform quintile scaling which mimics common portfolio sorting techniques (see \cite{GuKX2018}). After comparing different methods on the validation set, we settled for z-scoring as it maintains the distribution shape of the variables in contrast to uniform quantile scaling. It is important however to stress that different scaling methods may lead to different results and on our observations we found that some models might perform better using other scaling methods.
The mathematical formula corresponding to such transformation is as follows:
\begin{align*}
	x_{i, t}^{sc} = \frac{(x_{i, t} - \bar{x}_{., t})}{\sigma(x_{., t})},
\end{align*}
where $\bar{x}_{., t}$ and $\sigma(x_{., t}$ are the cross-sectional mean and standard deviation of the observations $(x_{1, t}, x_{N_t, t})$. Note that we proceed with the rescaling after winsorizing extreme values (beyond 3 inter-quartiles of the median since variables are not normally distributed). When omitting missing observations, the transformation standardizes all variables to have a mean 0 and unit variance. For the predicted variables, log excess returns and maximum drawdown, we would leave the variables unchanged.\\
\\
Before running the models, we need to deal with missing data. This issue is pervasive and important in machine learning applications. Removing stocks with missing information is not a solution as it will result in very few observations. Therefore, we rely on imputation methods. The most common imputation methods are mean, median, mode, and K-nearest neighbors. In our case, replacing missing values for categorical variables by their median and those in numerical variables by their mean gave better results on the validation set. Once again, the median and mean are taken over stocks on a given date.

\subsection{Performance and feature importance}\label{subsection:performancefeatureimportance}
Performance measures aim to evaluate a statistical model and are usually reported on out-of-sample (OOS) data, i.e., on data that are not used to calibrate the model. In the best-case scenario, a model's prediction should be as close as possible to the real outcome. A measure of the error between the forecast and the actual value can be used to interpret the results. Since regression models typically minimize the mean squared error, we display it as one of the performance measures. However, as an investor, the value is not necessarily informative and a mean squared error could be small but the trading strategy resulting from the prediction could still be bad. Therefore, we display the Kendall rank correlation, which measures the ordinal association between two quantities, as a second performance measure as it has better information on the performance of decile ranked portfolios\footnote{The coefficient of determination, or R-squared, which measures the proportion of the variance in the dependent variable that is predictable from the independent variable, is commonly used for regression analyses. Our reason for choosing an alternative measure is our focus on the ranking of stocks by their level of risk.}. \\
\\
We recall the formulas for the mean squared error (MSE) for our pooled data:
\begin{align*}
    \text{MSE} = \frac{1}{NT}\sum_{t\in \mathcal{T}}\sum_{i}^{N_t}(y_{i,t} - \hat{y}_{i,t})^2,
\end{align*}
where $NT=\sum_{t\in \mathcal{T}}N_t$ denotes the number of observations in the test set. For the MSE, the model is better when the MSE value is smaller.\\
\\
The Kendall rank correlation ($\tau$) for a sample of size $N$ is given by the following formula:
\begin{align*}
\tau = \frac{\text{(number of concordant pairs) - (number of discordant pairs)}}{\binom{N}{2}}.
\end{align*}
We apply the formula for each prediction date $t$:
\begin{align*}
\tau_t = \frac{1}{2N_t(N_t - 1)}\sum_{i < j}{\text{sgn}({y}_{i, t} - {y}_{j, t})\times \text{sgn}(\hat{y}_{i, t} - \hat{y}_{j, t})},
\end{align*}
where $N_t$ is the number of stocks available at $t$.\\
\\
To calculate the performance measure over the entire test period, we average the time series of Kendall $\tau$ over the test set $\mathcal{T}$:
\begin{align*}
\tau = \frac{1}{T}\sum_{t \in \mathcal{T}}\tau_t.
\end{align*}
\noindent In theory, the overall Kendall correlation ranges from -1 to 1. A value of 1 means that the predicted MDD ranks the stocks exactly as realized MDD does. A correlation close to 0 can be interpreted as no relationship between the prediction and realization, while a negative correlation means a discordance between the prediction and realized MDD rankings.\\
\\
Note that any other correlation measure (e.g., Spearman or Pearson) works as well. The choice of the Kendall rank correlation is mainly due to its simplicity to interpret.
\\
\color{black}
The idea of feature importance is to measure how much a performance metric decreases when a feature is not available. A potential way to measure feature importance would be to remove the feature from the data set, re-train the model with the other estimators, and measure the change in performance. On the one hand, this solution requires re-training and can be computationally intensive. On the other hand, it shows what feature is important within the data set rather than within a given trained model\footnote{Unlike simple models like linear regressions, where the ``beta'' coefficients and t-statistic give information on the importance of a regressor, more complex models, like neural networks or random forest, don't have such parameters and thus require to resort to more advanced feature importance analyses.}.\\
\\
One goal of the paper is to find characteristics that enable one to predict future stock risk and performance to a certain extent. While asset pricing theory relies on linear regression models and their coefficients. Our multiple models, some of which are nonparametric, require a unified framework. Luckily, a large range of literature on machine learning tackles this question. \cite{RibeiroTG2016} propose the ``local interpretable model-agnostic explanations". Their method relies on local surrogate models that are trained to approximate the predictions of the underlying black box model. A surrogate model can be any interpretable model such as linear regression or a decision tree. \cite{WeiLS2015} review and compare a large set of methods for variance importance models (VIM) including difference-based VIMs, hypothesis test techniques, and variance-based VIMs. The intuition behind these methods is that the more a model's decision criteria depend on a feature, the more the predictions change as a function of perturbing this feature. \cite{Breiman2001} and \cite{AltmannTSL2010} suggest a permutation importance method, which measures how a score decreases when the empirical values of a feature are replaced by  random noise drawn from the distribution of feature values. In their analysis of the predictability of monthly returns, \cite{GuKX2018} measures the importance of a variable by the reduction in $R^2$ obtained by setting all the values of the selected variable to zero and keeping the values of the other variables fixed. \\
\\
In choosing the best method for the application, one needs to consider a trade-off between efficacy and application. Permutation methods can be time-consuming for large data sets and like other methods, ignore the correlation between the variables for certain models. Since we are not interested in inference in this paper, we opt for the same approach as \cite{GuKX2018} by setting one column to zero and measuring the change in the Kendall correlation when doing so.\\
\\
For the second part of the analysis where ESG data are incorporated, we use a more novel approach in addition to the feature importance from the first part. The new approach is called Shapley additive explanations (SHAP)\cite{LundbergL17}. SHAP stems from game theory and assigns importance to variables based on their contribution to improving the model from different \textit{collaborative combinations}. The complexity of Shapley value is NP-hard which makes applying it to all the models very time-consuming. There are approximations that are polynomial in time for tree-based models. Therefore, we focus on XGBoost to conduct the analysis using Shapley values.

\section{Empirical Study}
We collect monthly data from WRDS, OWL Analytics, and TruValue Labs for stocks listed in the Russel 3000 index. Our data set starts in January 1970 and ends in December 2019. We account for 119 non-ESG characteristics in total and 20 ESG-related variables, 16 from OWL Analytics and 4 from TruValue Labs.\\
\\
The regression is performed by considering the log excess returns and maximum drawdown  (without any transformation as opposed to the explanatory variables) as the dependent variable and the matrix of rescaled firm characteristics as the predictors. The models trained along with their parameters are given in Table \ref{tab:hyperparams_logER} and \ref{tab:hyperparams_mdd}. We compare the results based on out-of-sample performance

\subsection{A first analysis without ESG data}
Because ESG data begin in 2007, we omit ESG scores for a first analysis in which we test the models using the other 111 variables. This enables us to carry out the analysis on a larger test set that includes the 2008 crisis. The training set starts in January 1970 and ends in December 2000, while the test set starts in January 2001 and ends in December 2019.
\subsubsection{The cross-section of returns}
The models were trained after tuning the hyperparameters given in Table \ref{tab:hyperparams_logER}.
\begin{table}[H]
\centering
	\caption{Hyperparameters for training different models of log excess returns}
\resizebox{0.7\width}{!}{\begin{tabular}{l|llllllll}
\toprule
{} & Lasso &    Ridge &   ENet &  PLS &  PCR &     RF &           XGBoost &  MLP \\
\midrule
Penalization ($\alpha$)            &  $10^{-2}$ &  $10^{4}$ &  $10^{-3}$ &      &      &        &               $10^{-1}$ &  $10^{-1}$ \\
Number of components     &       &          &        &  6 &  8 &        &                   &      \\
Number of estimators     &       &          &        &      &      &  300 &              $10^{3}$ &      \\
Maximum depth        &       &          &        &      &      &   10 &                 3 &      \\
Learning rate    &       &          &        &      &      &        &              $5\times 10^{-2}$ &      \\
Subsample ratio of columns by tree &       &          &        &      &      &        &               0.3 &      \\
Number of neurons        &       &          &        &      &      &        &                   &  3 \\
Number of units           &       &          &        &      &      &        &                   &  8 \\
\bottomrule
\end{tabular}
}
\label{tab:hyperparams_logER}
\end{table}
\subsubsection*{Performance}
\begin{figure}[h]
	\caption{Mean squared error and Kendall correlation (logER)}
	\caption*{The table and barplot report the overall MSE and the smallest, resp. largest 1000 stocks, and cover the nine models, i.e., linear regression (OLS), penalized regressions (Lasso, Ridge, and ENet), dimension reduction methods (PCR and PLS), tree-based models (RF and XGBoost), and multi-layer perceptron (MLP). The test period is from 2001-01 until 2019-12.}
    \centering
    \begin{subfigure}{1\textwidth}
		\caption{Table of out-of-sample overall MSE}
		\begin{tabular}{llllllllll}
\toprule
{} &   OLS & Lasso & Ridge &  ENet &   PCR &   PLS &    RF & XGBoost &   MLP \\
\midrule
All companies                & 0.249 & 0.249 & 0.249 & 0.249 & 0.250 & 0.249 & 0.251 &   0.251 & 0.244 \\
Top 1000 companies (size)    & 0.151 & 0.150 & 0.151 & 0.150 & 0.150 & 0.151 & 0.151 &   0.153 & 0.146 \\
Bottom 1000 companies (size) & 0.366 & 0.368 & 0.365 & 0.365 & 0.367 & 0.365 & 0.370 &   0.367 & 0.360 \\
\bottomrule
\end{tabular}

    \end{subfigure}
    \begin{subfigure}{1\textwidth}
		\caption{Table of out-of-sample average Kendall correlation}
		\begin{tabular}{lrrrrrrrrr}
\toprule
{} &   OLS &  Lasso &  Ridge &  ENet &   PCR &   PLS &    RF &  XGBoost &   MLP \\
\midrule
All companies                &  9.09 &   9.45 &   9.12 &  9.26 &  8.97 &  9.16 &  8.57 &     9.56 &  9.17 \\
Top 1000 companies (size)    &  5.94 &   5.92 &   5.97 &  5.97 &  5.48 &  5.92 &  4.79 &     6.97 &  6.10 \\
Bottom 1000 companies (size) & 11.36 &  11.59 &  11.34 & 11.48 & 10.97 & 11.46 & 10.98 &    11.56 & 11.20 \\
\bottomrule
\end{tabular}

    \end{subfigure}
    \begin{subfigure}{1\textwidth}
        \caption{Barplot of out-of-sample overall MSE (left) and Kendall correlation (right)}
        \centering
		\includegraphics[height=7cm, width=8.3cm]{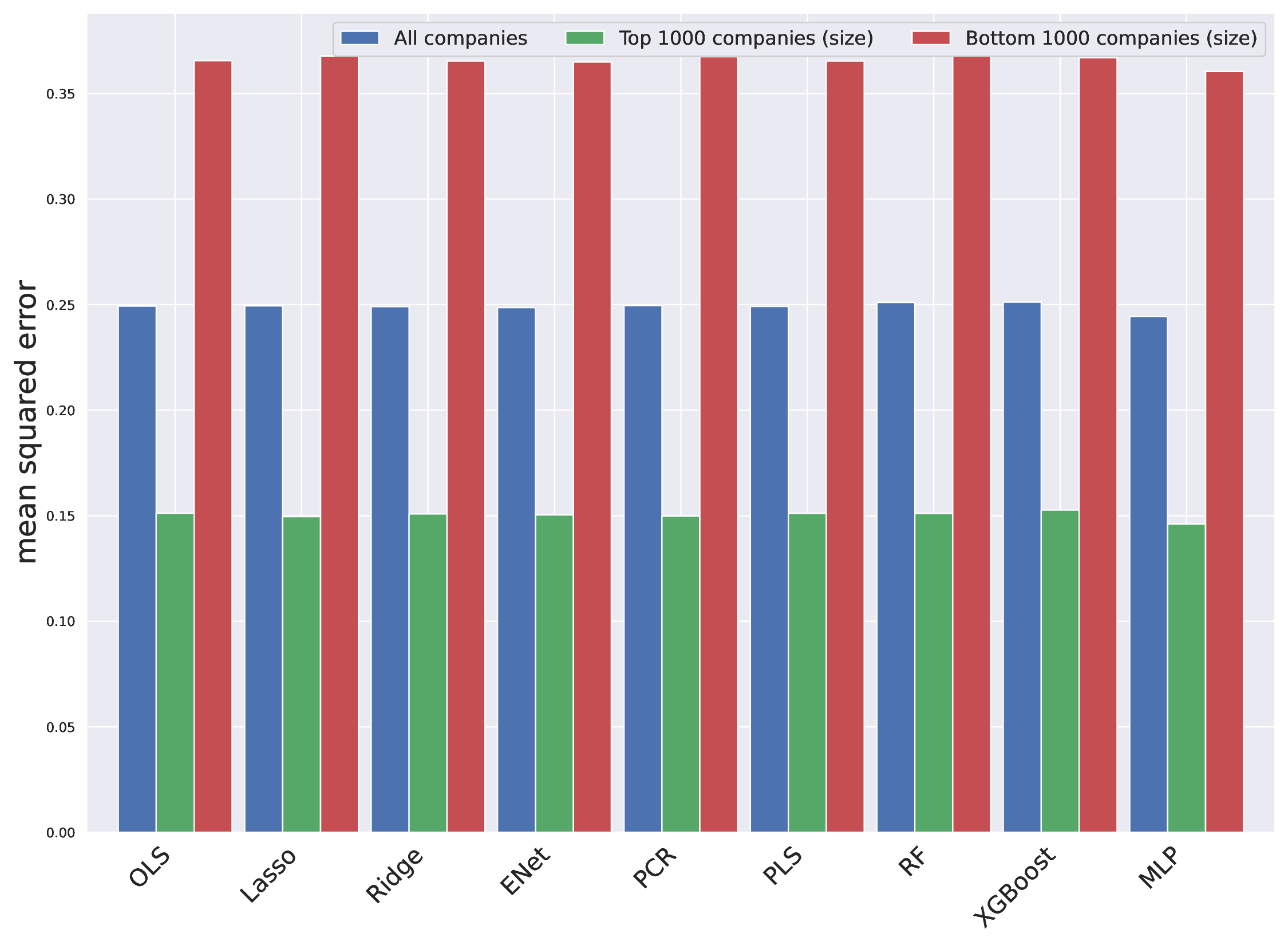}
		\includegraphics[height=7cm, width=8.3cm]{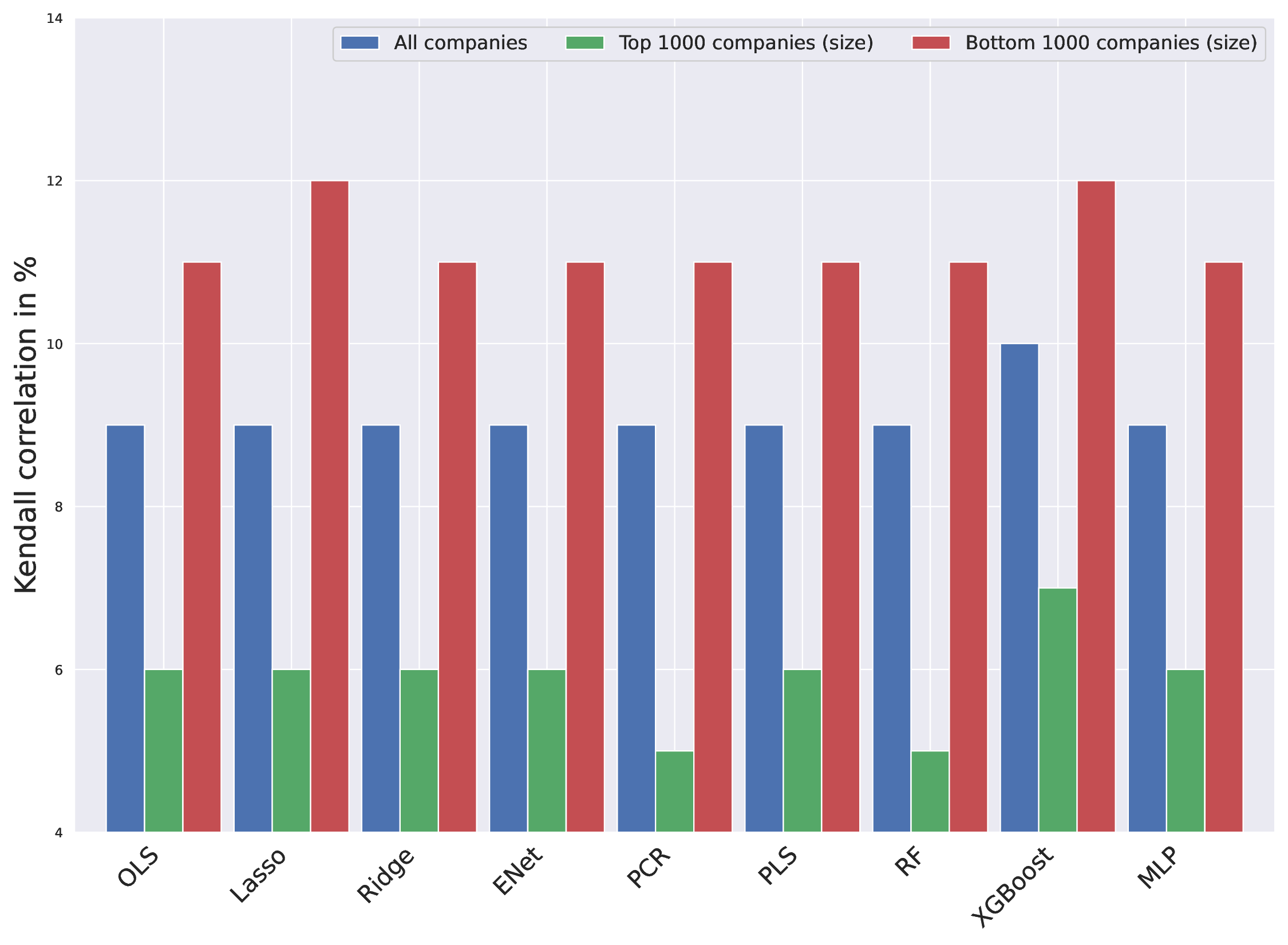}
    \end{subfigure}    
\label{fig:tau_noESG_ret}
\end{figure}

\begin{figure}[H]
\caption{The evolution of Kendall correlation in time between 2000-01 and 2019-12 ( (logER)}
\centering
\includegraphics[scale=0.3]{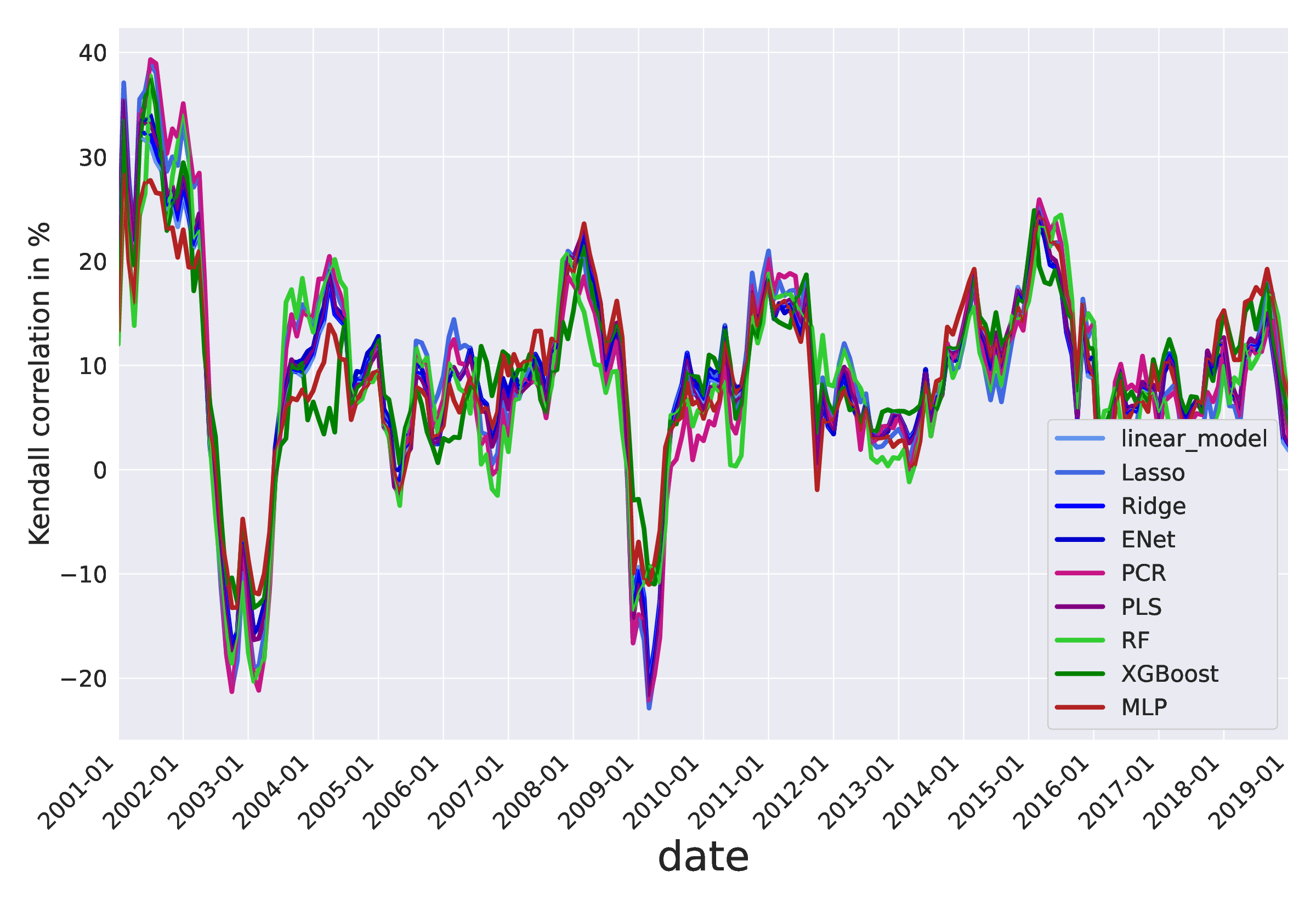}
\label{fig:models_vs_time_noESG_ret}
\end{figure}

In Figure \ref{fig:tau_noESG_ret}, we report the MSE and average Kendall correlation for all the stocks, and the largest and smallest 1000 stocks by market value. XGBoost achieved the best correlation performance ($9.56\%$), while MLP achieved the smallest MSE (0.244). RF and PCR are slightly below other models. Overall, the values of the performance measures are very close from one model to another.
\noindent The previous results are aggregated over the entire test period, and they do not inform us about when the models performed poorly and when they gave a sound estimation. Also, because the MSE is less informative regarding the impact on investment strategies, we focus on the cross-sectional correlation for each date of the test period, leading to a time series of out-of-sample Kendall correlation. The results are shown in Figure \ref{fig:models_vs_time_noESG_ret}. All models performed poorly during 2003 and the 2008 financial crisis without exception. The cross-sectional Kendall correlation reached its minimum (around $-20\%$) in both periods, which suggests that the models would have failed to provide a useful signal for investors. To inspect this claim further, we divide our stocks for each date into five quintiles based on their predicted maximum drawdown. We then plot the empirical distributions of the stocks' realized drawdown.
\begin{figure}[H]
\caption{Realized logER distribution for predicted logER-based quantiles}
\caption*{The quintiles $q1, ..., q5$ are formed using the empirical distribution of predicted logER for each of the two dates 2008-01-01 and 2016-01-01. The distribution of realized log excess return is then displayed for each quintile over these periods for the MLP method.}
\centering
\includegraphics[scale=0.2]{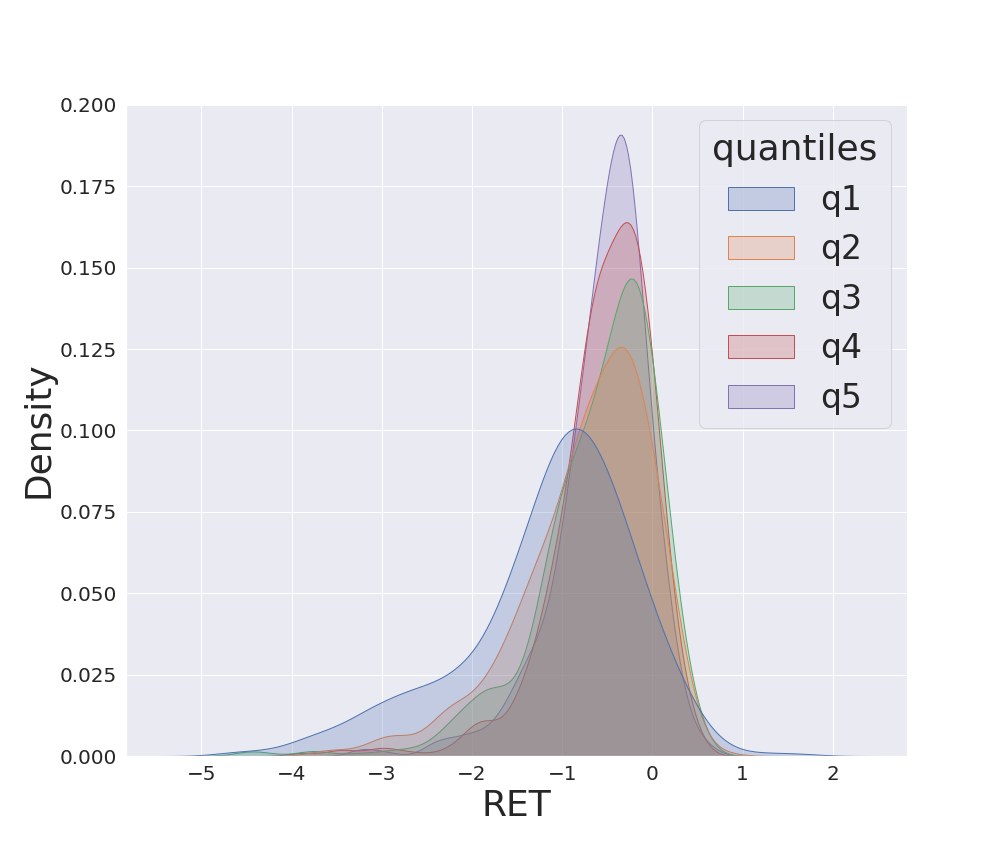}
\includegraphics[scale=0.2]{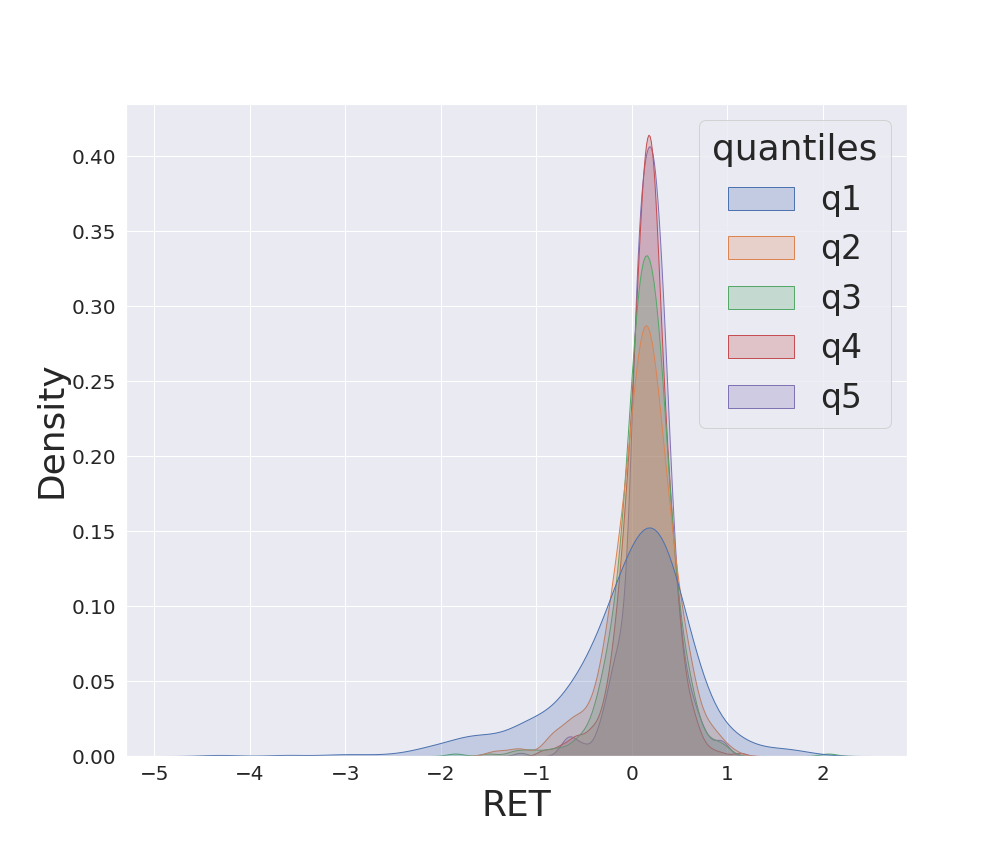}
\label{fig:quantiles_noESG_ret}
\end{figure}
\noindent Figure \ref{fig:quantiles_noESG_ret} displays the relationship between predicted and realized log excess returns over two year-long periods. For each period, stocks are sorted into five quintiles on the basis of predicted log excess returns, and the distributions of realized log excess returns are plotted. The first period (between 2008-01-01 and 2008-12-31) includes the financial crisis, while calmer market conditions prevailed during the second period (between 2016-01-01 and 2015-12-31). During the calmer period, the densities of different quintiles are more distinct. During the crisis, the bellies of the distributions tended to overlap with an average that decreases as the quintiles increase suggesting.

\color{black}
\subsubsection*{Feature importance}
\begin{figure}[H]
\caption{Feature importance for different models  (logER)}
\caption*{The frigure (heatmap) reports feature importance for the features that don't deteriorate the prediction over the test set. 
The results are reported for each of the nine models over the out-of-sample period between 2001-01 and 2019-12.
}
\centering
\includegraphics[height=16cm, width=13cm]{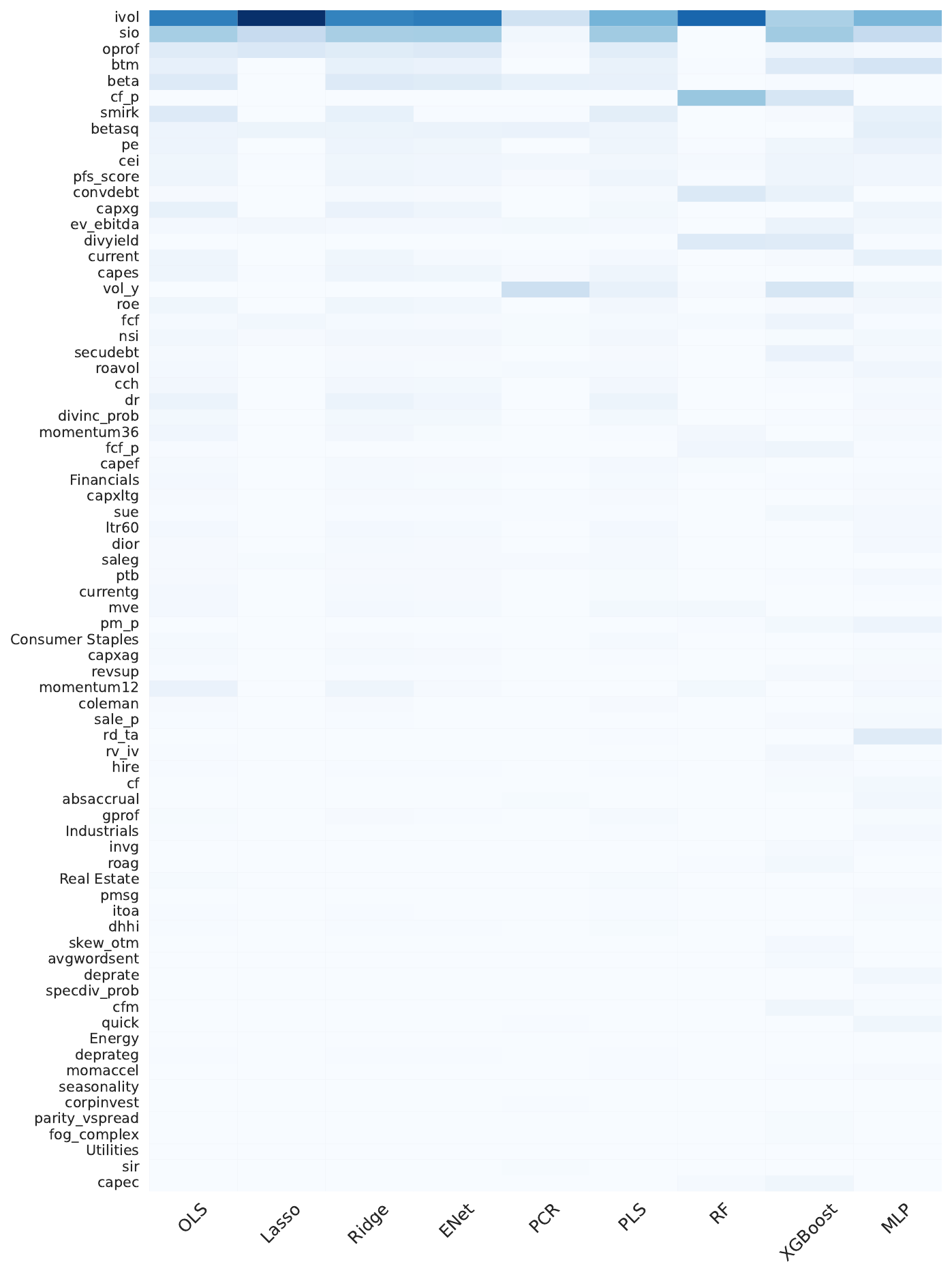}
\label{fig:feature_importance_noESG_ret}
\end{figure}

To explore the predictive power of different firm characteristics, we calculate a feature importance score for all the variables over the test set. We display variables that don't decrease the performance which accounts for 72. Reported in Figure \ref{fig:feature_importance_noESG_ret}, these results show that, volatility-based and liquidity variables, i.e., idiosyncratic volatility (ivol), short-interest (sio), operating profitability (oprof), book-to-market (btm) and beta. Momentum (momentum36) and size (mve) are also among the important features. Overall, the models agree on several variables and disagree on others. We find without surprise that volatility and idiosyncratic volatility outclass the other variables.
\subsubsection*{Performance of decile portfolios}
\begin{figure}[H]
\caption{The cumulative returns of equally-weighted quintile portfolios over time b/w 2001-06 and 2019-12 for XGBoost  (logER).}
\caption*{The decile portfolios are equally-weighted portfolios formed using predicted log excess return deciles (from low to high) and rebalanced annually. The market portfolio (mkt) is an equally-weighted portfolio with annual rebalancing frequency.}
\centering
\includegraphics[height=9cm, width=14cm]{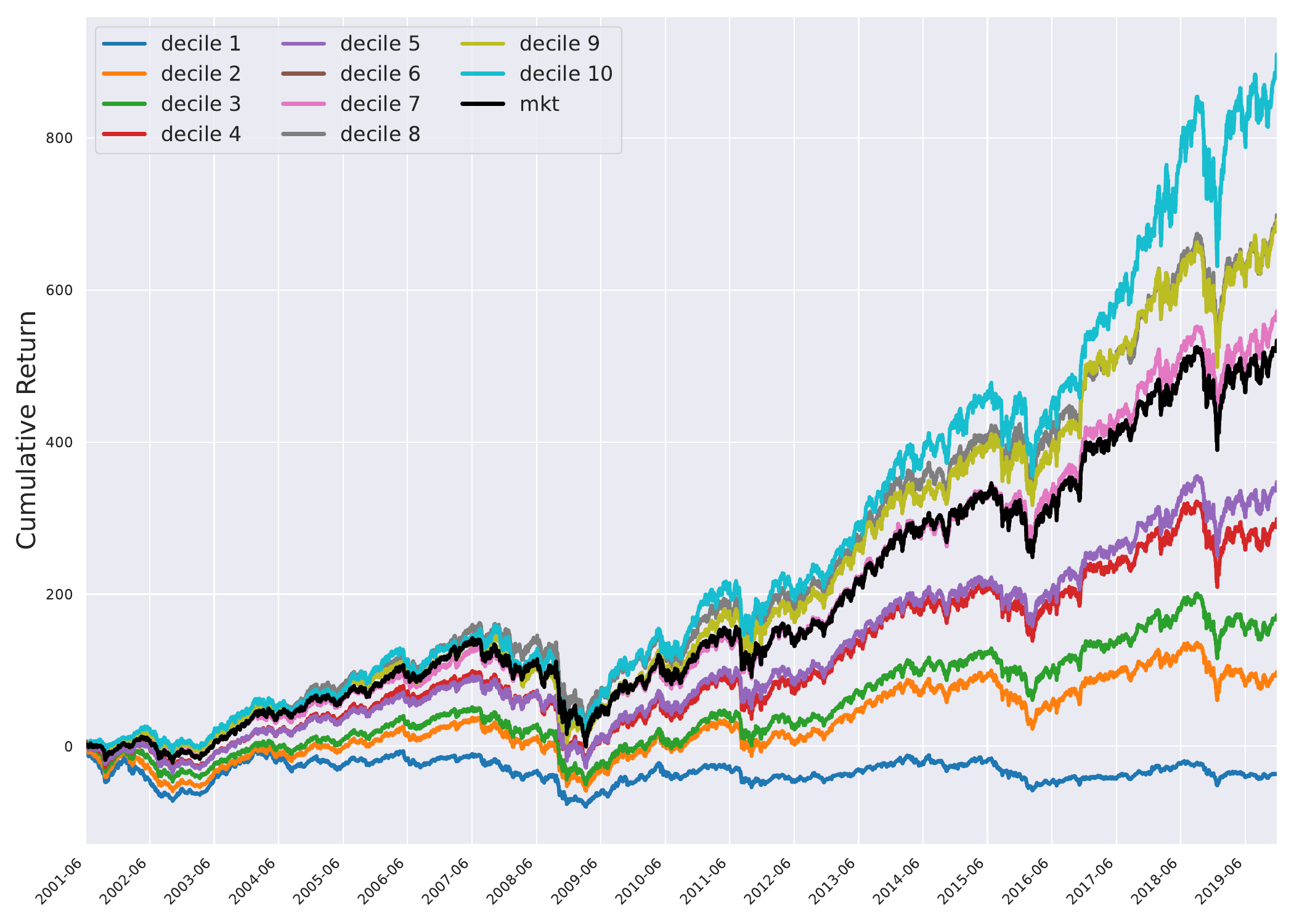}
\label{fig:quantile_noESG_ret}
\end{figure}

We investigate the performance of quintile portfolios based on predicted log excess returns with annual rebalancing. We form portfolios based on deciles of the predicted quantity each first of June of each year (06-01)\footnote{We adopt this convention to be consistent with common practice relying on new companies information being published around that period, and also to mitigate market conditions around the new year.}. The portfolio is held for one year and rebalanced on the next first of June based on new stock deciles. We display in Figure \ref{fig:quantile_noESG_ret} the cumulative returns of the equally-weighted portfolios and report in Table \ref{tab:table_perf_noESG_ret} some statistics for these portfolios. We report the results for XGBoost but other models exhibit similar results\footnote{We also performed the analysis on market value-weighted portfolios, and the results were consistent.}. Sharpe and Calmar ratios are mostly increasing as a function of the quintiles (as the predicted logER increases) which is desirable for a portfolio manager.
\begin{table}[h]
\centering
	\caption{Portfolio performance metrics for equally-weighted portfolios for XGBoost  (logER)}
	\caption*{The average of annualized return (RET), volatility (VOL), maximum drawdown (MDD), and Sharpe and Calmar ratios for the period 2001-2019 for decile portfolios and the market (mkt). Portfolios are equally-weighted and the predictions are from the XGBoost model.}
\begin{tabular}{l|rrrrr}
\toprule
{} &    RET &    VOL &    MDD &    Sharpe Ratio &    Calmar Ratio \\
\midrule
mkt       &  11.04 &  20.16 &  18.43 &  0.49 &  1.04 \\
\midrule
decile 1  &   5.87 &  26.27 &  29.53 &  0.13 &  0.58 \\
decile 2  &   9.08 &  23.13 &  23.28 &  0.35 &  0.87 \\
decile 3  &   9.09 &  21.57 &  20.68 &  0.38 &  0.85 \\
decile 4  &  10.30 &  20.49 &  18.26 &  0.47 &  1.04 \\
decile 5  &  11.01 &  19.82 &  17.39 &  0.50 &  1.08 \\
decile 6  &  12.34 &  19.44 &  16.46 &  0.60 &  1.21 \\
decile 7  &  12.29 &  18.74 &  15.55 &  0.60 &  1.18 \\
decile 8  &  13.08 &  18.59 &  15.19 &  0.65 &  1.26 \\
decile 9  &  13.54 &  18.70 &  15.76 &  0.67 &  1.33 \\
decile 10 &  14.69 &  19.88 &  16.55 &  0.68 &  1.36 \\
\bottomrule
\end{tabular}
 
\label{tab:table_perf_noESG_ret}
\end{table}
\noindent Figure \ref{fig:quantile_noESG_ret} shows how the high logER quintile (decile 10) outperformed other deciles. We also note that the low logER quintile (decile 1) is the one with the lowest Sharpe and Calmar ratios.
\newpage
\subsubsection{The cross-section of maximum drawdown}
In the following section, we repeat the same set of analyses to maximum drawdown using the same periods, models, and performance measures. The models were trained after tuning the hyperparameters given in Table \ref{tab:hyperparams_mdd}.
\begin{figure}[H]
	\caption{Out-of-sample MSE and average Kendall correlation (MDD)}
	\caption*{The table and barplot report the overall MSE and the smallest, resp. largest 1000 stocks, and cover the nine models, i.e., linear regression (OLS), penalized regressions (Lasso, Ridge, and ENet), dimension reduction methods (PCR and PLS), tree-based models (RF and XGBoost), and multi-layer perceptron (MLP). The test period is from 2001-01 until 2019-12.}
    \centering
    \begin{subfigure}{1\textwidth}
		\caption{Table of out-of-sample overall MSE}
		\begin{tabular}{llllllllll}
\toprule
{} &   OLS & Lasso & Ridge &  ENet &   PCR &   PLS &    RF & XGBoost &   MLP \\
\midrule
All companies                & 0.028 & 0.028 & 0.028 & 0.028 & 0.029 & 0.028 & 0.030 &   0.028 & 0.027 \\
Top 1000 companies (size)    & 0.024 & 0.024 & 0.024 & 0.024 & 0.025 & 0.024 & 0.025 &   0.025 & 0.024 \\
Bottom 1000 companies (size) & 0.032 & 0.032 & 0.032 & 0.032 & 0.033 & 0.033 & 0.035 &   0.031 & 0.031 \\
\bottomrule
\end{tabular}

    \end{subfigure}
    \begin{subfigure}{1\textwidth}
		\caption{Table of out-of-sample overall Kendall correlation}
		\begin{tabular}{lrrrrrrrrr}
\toprule
{} &   OLS &  Lasso &  Ridge &  ENet &   PCR &   PLS &    RF &  XGBoost &   MLP \\
\midrule
All companies                & 47.14 &  47.11 &  47.13 & 47.13 & 45.26 & 47.16 & 44.41 &    47.71 & 47.01 \\
Top 1000 companies (size)    & 39.96 &  39.83 &  39.95 & 39.90 & 36.79 & 39.94 & 35.29 &    40.31 & 39.64 \\
Bottom 1000 companies (size) & 45.39 &  45.40 &  45.39 & 45.40 & 43.70 & 45.45 & 43.12 &    46.47 & 45.47 \\
\bottomrule
\end{tabular}

    \end{subfigure}
    \begin{subfigure}{1\textwidth}
        \caption{Barplot of out-of-sample overall MSE (left) and Kendall correlation (right)}
        \centering
		\includegraphics[height=6cm, width=8.3cm]{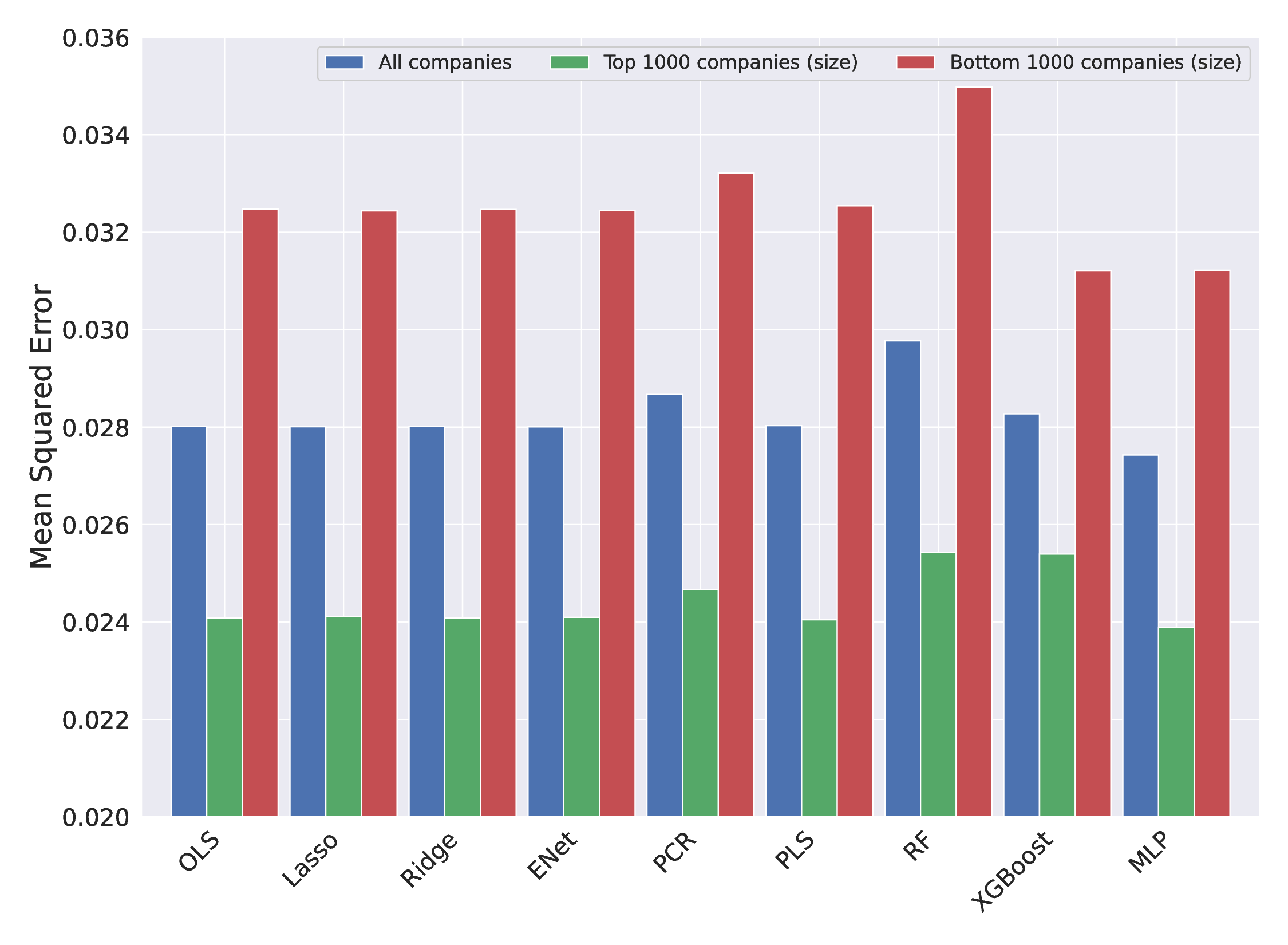}
		\includegraphics[height=6cm, width=8.3cm]{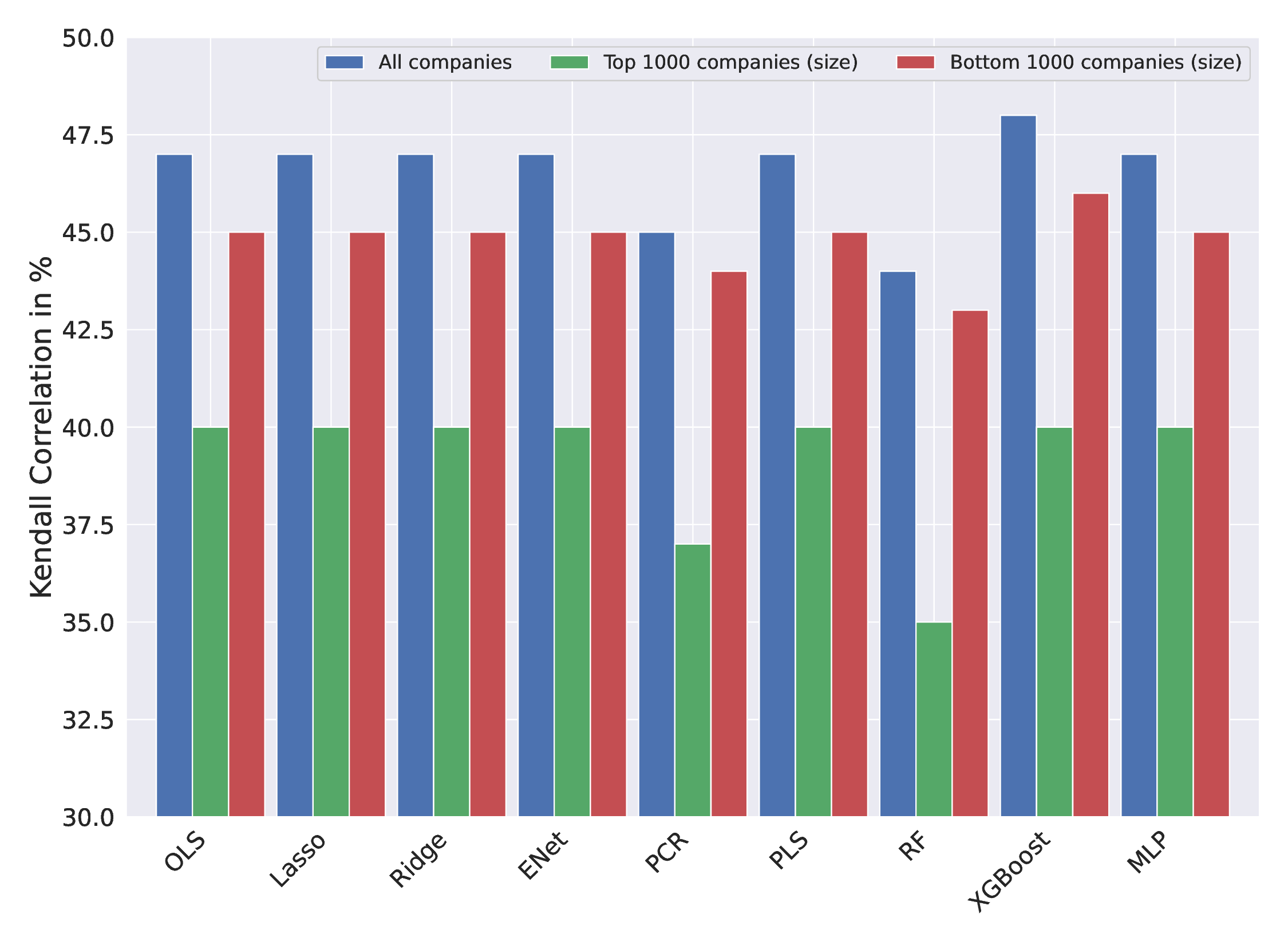}
    \end{subfigure}    
\label{fig:tau_noESG_mdd}
\end{figure}
\begin{table}[H]
\centering
	\caption{Hyperparameters for training different models of maximum}
\resizebox{0.7\width}{!}{\begin{tabular}{l|llllllll}
\toprule
{} &   Lasso &   Ridge &    ENet &  PLS &   PCR &     RF &           XGBoost &   MLP \\
\midrule
Penalization ($\alpha$)            &  $10^{-4}$ &  $10^{3}$ &  $10^{-4}$ &      &       &        &               $10^{-1}$ &   $10^{-1}$ \\
Number of components     &         &         &         &  8 &  35 &        &                   &       \\
Number of estimators     &         &         &         &      &       &  300 &               500 &       \\
Maximum depth        &         &         &         &      &       &   10 &                10 &       \\
Learning rate    &         &         &         &      &       &        &              $10^{-2}$ &       \\
Subsample ratio of columns by tree &         &         &         &      &       &        &               0.3 &       \\
Number of neurons        &         &         &         &      &       &        &                   &   5 \\
Number of units           &         &         &         &      &       &        &                   &  64 \\
\bottomrule
\end{tabular}
}
\label{tab:hyperparams_mdd}
\end{table}
\newpage
\subsubsection*{Performance}
\begin{figure}[H]
\caption{The evolution of Kendall correlation in time between 2001-01 and 2019-12 (MDD)}
\centering
\includegraphics[scale=0.3]{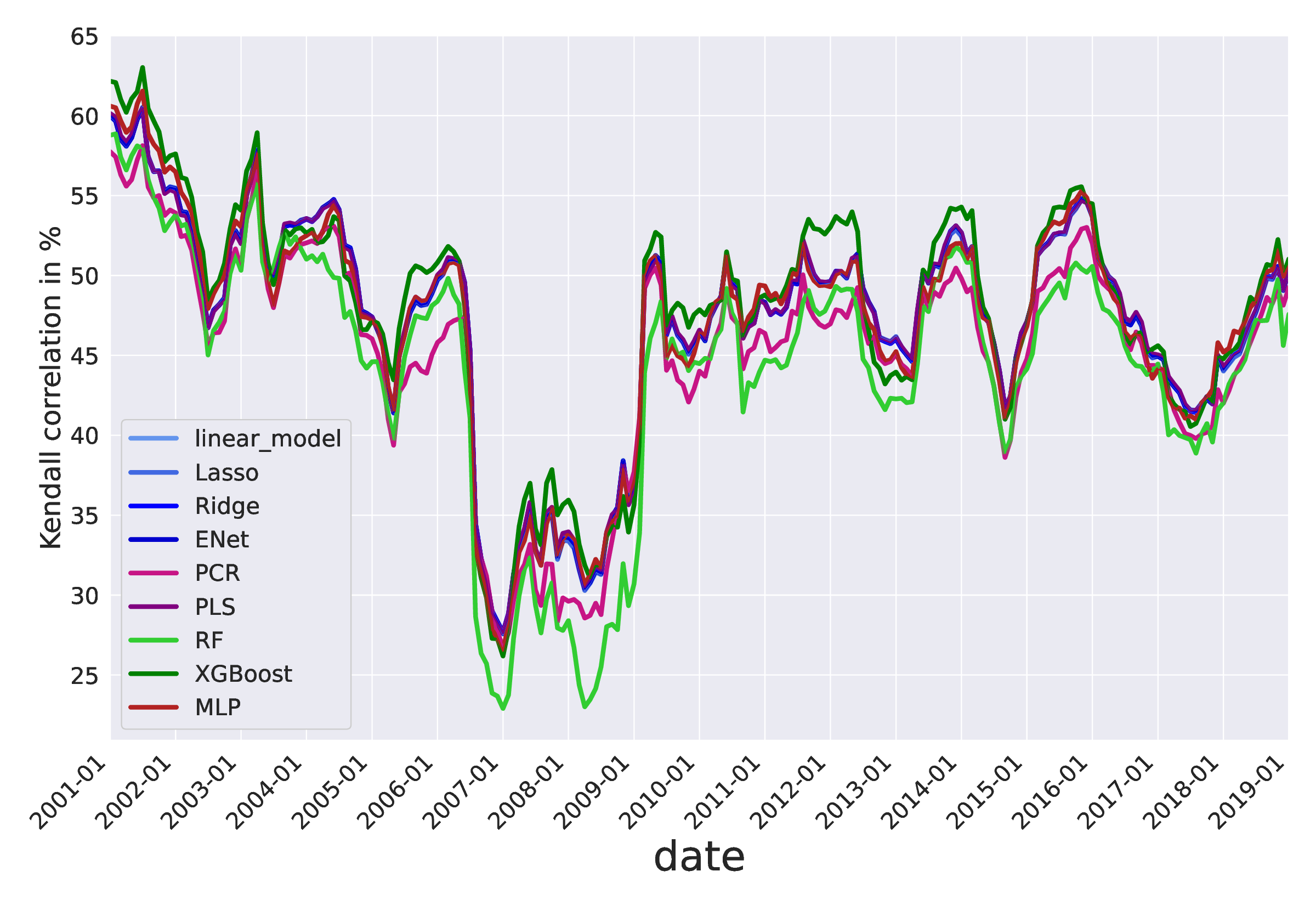}
\label{fig:models_vs_time_noESG_mdd}
\end{figure}

Figure \ref{fig:tau_noESG_mdd} reports the MSE and average Kendall correlation. The best performance was achieved by XGBoost ($47.71\%$ correlation and 0.028 MSE), followed by linear regression (OLS) ($47.14\%$ and 0.028 MSE). Random forest and principal component regression are again slightly underperforming the other models ($44.41\%$ and $45.26\%$ respectively). Overall, the values of the performance measures are very close from one model to another.
\noindent We observe again the time series of Kendall correlation for the period 2001-2019. Those results are shown in Figure \ref{fig:models_vs_time_noESG_mdd}. Without exception, all models performed poorly during the 2008 financial crisis between 2008 and mid-2009. The cross-sectional Kendall correlation reached its minimum but remained positive (slightly above $25\%$). This value indicates that, despite the poor performance, the concordance between the prediction and realized maximum drawdown is persistent throughout the period.
\begin{figure}[H]
\caption{Realized MDD distribution for predicted MDD-based quantiles}
\caption*{The quintiles $q1, ..., q5$ are formed using the empirical distribution of predicted MDD for each of the two dates 2008-01-01 (left) and 2016-01-01 (right). The distribution of realized MDD is then displayed for each quintile over these periods for the MLP method.}
\centering
\includegraphics[scale=0.2]{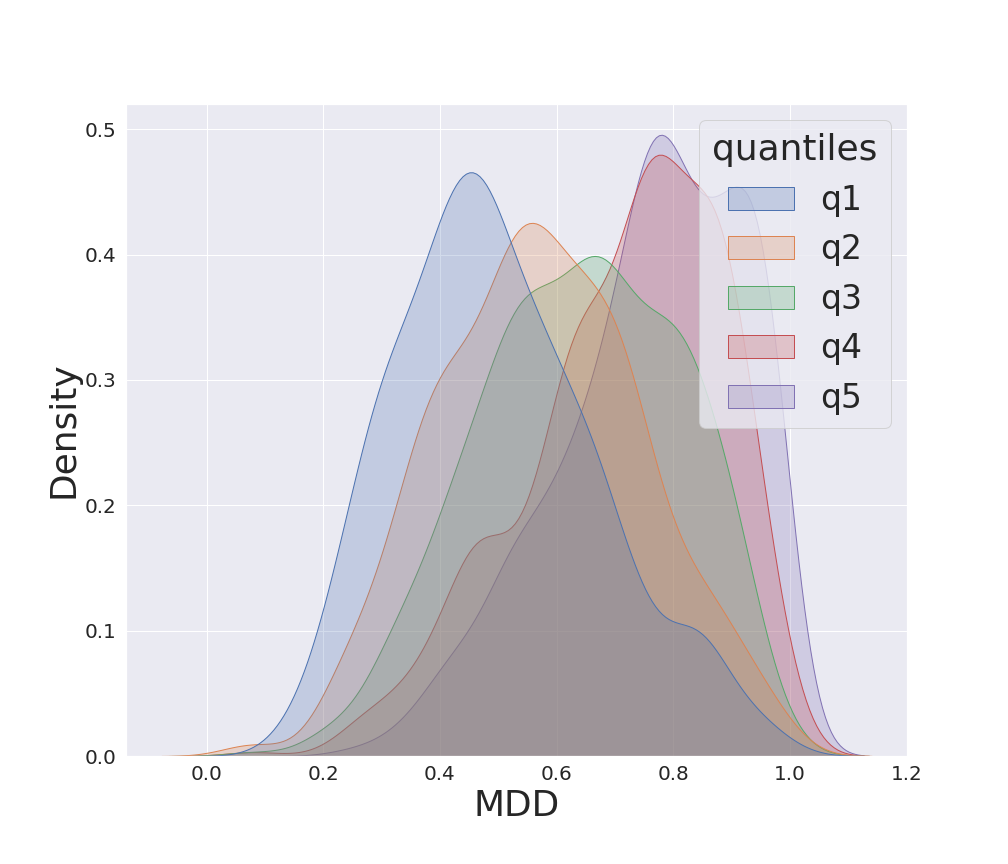}
\includegraphics[scale=0.2]{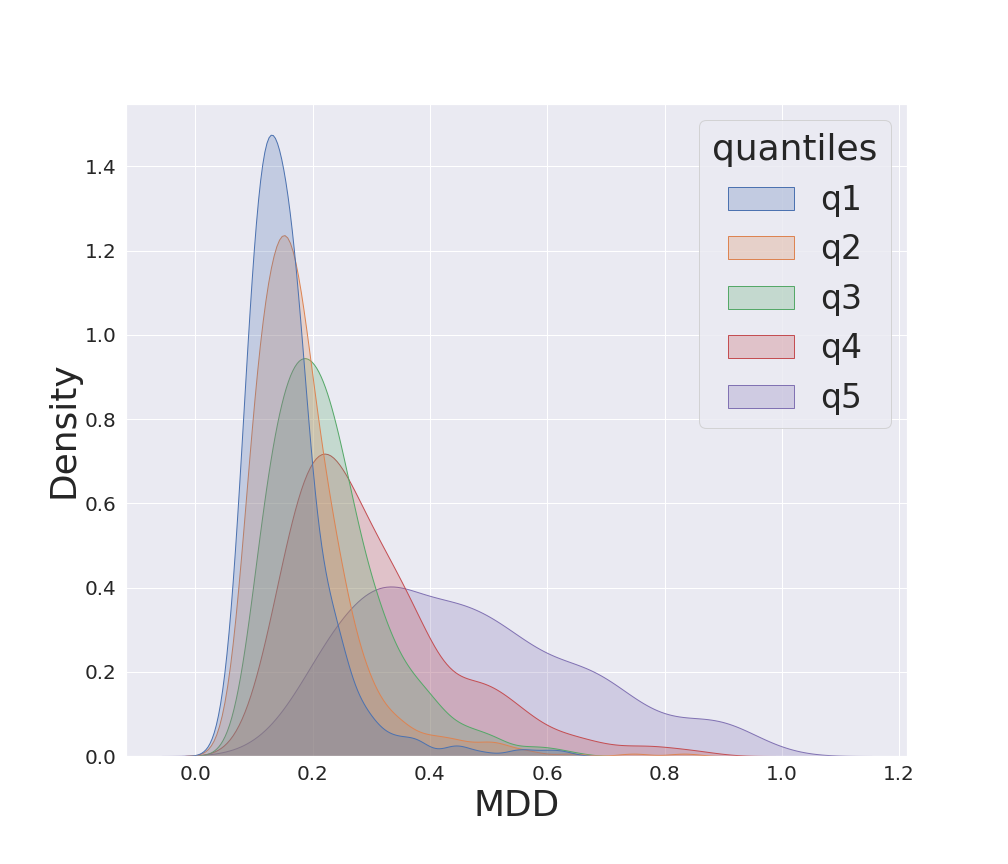}
\label{fig:quantiles_noESG_mdd}
\end{figure}
\newpage
\noindent Figure \ref{fig:quantiles_noESG_mdd} displays the relationship between predicted and realized maximum drawdown over two year-long periods. For each period, stocks are sorted into five quintiles on the basis of predicted maximum drawdown, and the distributions of realized maximum drawdown are plotted. The first period (between 2008-01-01 and 2008-12-31) includes the financial crisis, while calmer market conditions prevailed during the second period (between 2016-01-01 and 2016-12-31). During the calmer period, the densities of different quintiles are more distinct. During the crisis, the middle quantiles tended to overlap while extreme quintiles remained well distinguished.
\subsubsection*{Feature importance}
\begin{figure}[h]
\caption{Feature importance for different models  (MDD)}
\caption*{The figure (heatmap) reports feature importance defined by the change in Kendall correlation  when setting the variable to zero. The results are reported for each of the nine models over the out-of-sample period between 2001-01 and 2019-12.
}
\centering
\includegraphics[height=15cm, width=12cm]{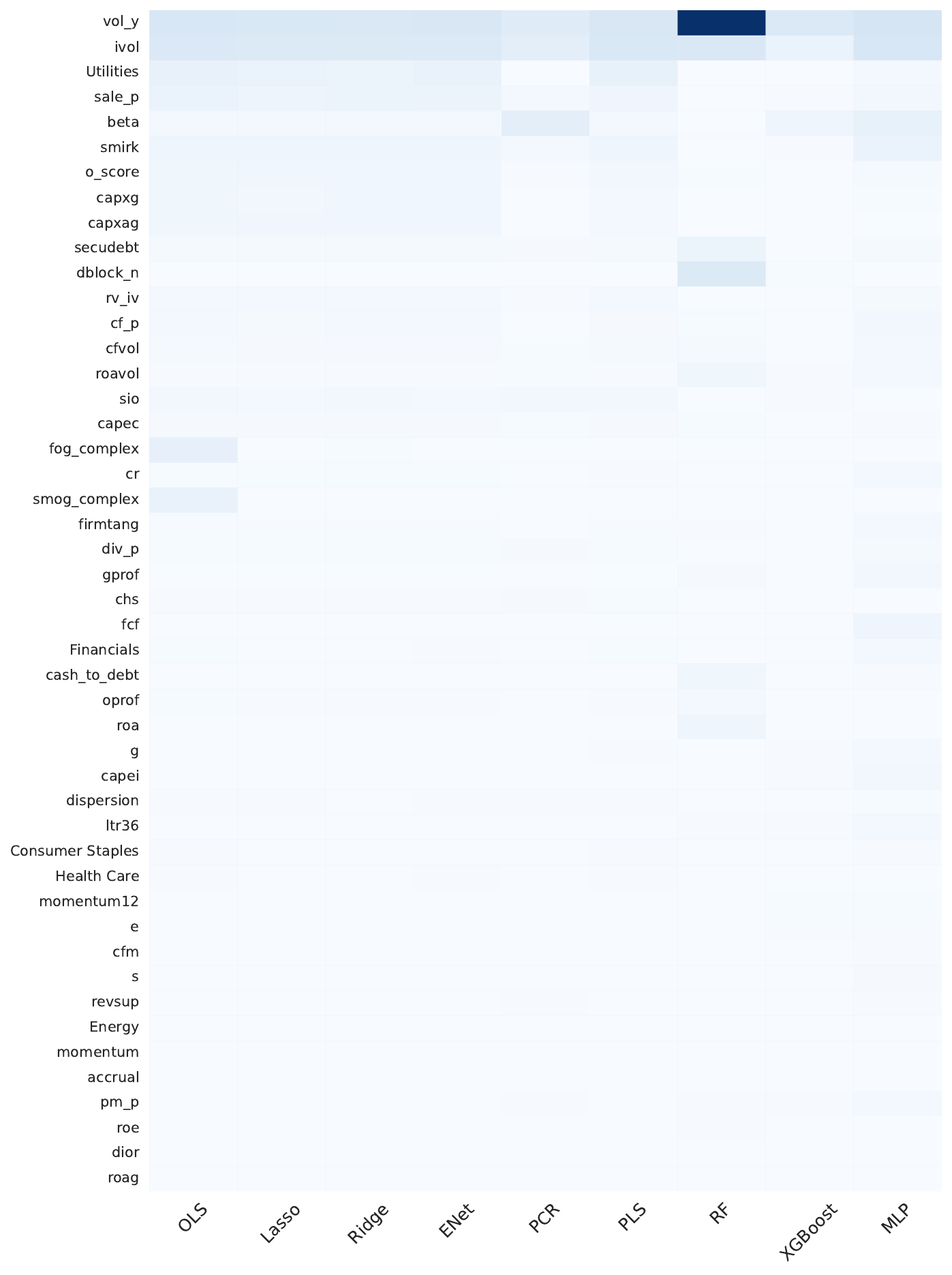}
\label{fig:feature_importance_noESG_mdd}
\end{figure}

We explore again the power of different firm characteristics and their contribution to the performance of the models out-of-sample. We show the ranking of the predictors that contributed to the performance of the models on average. Reported in Figure \ref{fig:feature_importance_noESG_mdd}, these results show that, volatility-based and liquidity variables, i.e., idiosyncratic volatility (ivol), realized volatility (vol\_y), and beta are the top predictors. We also find among those variables short interest and option-based variables (sio, smirk). Overall, the models exhibit the same predominant features but also have differences among other weaker predictors. We find without surprise that volatility and idiosyncratic volatility outclass the other variables.
\subsubsection*{Performance of decile portfolios}
\begin{figure}[ht]
\caption{The cumulative returns of equally-weighted decile portfolios over time b/w 2001 and 2019 for XGBoost (MDD)}
\caption*{The decile portfolios are formed using predicted maximum drawdown deciles (from low to high) and rebalanced at the beginning of each year and held until the end of the year. The market portfolio (mkt) is an equally-weighted portfolio with yearly rebalancing frequency.}
\centering
\includegraphics[height=8cm, width=12cm]{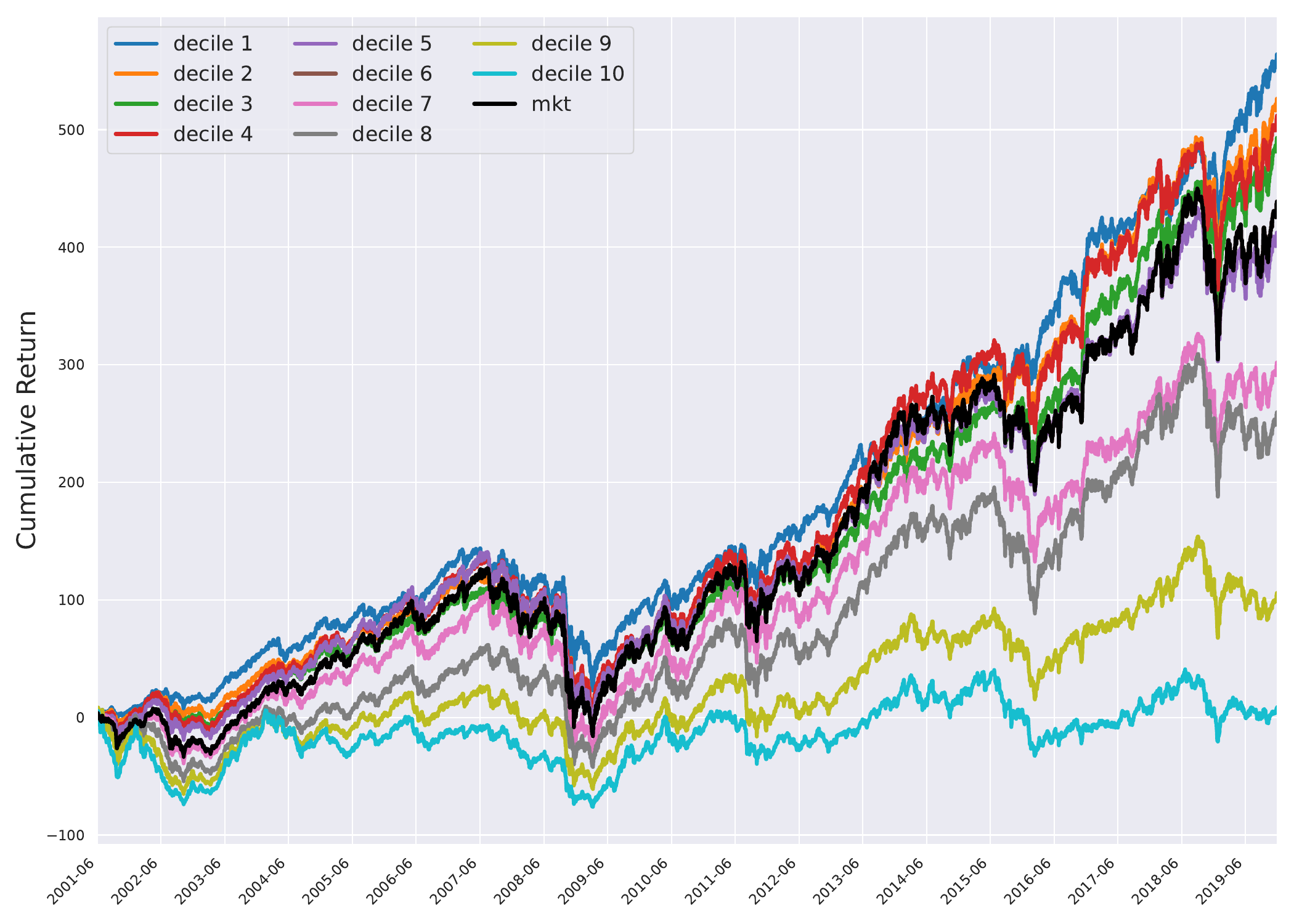}
\label{fig:quantile_noESG}
\end{figure}

We investigate the performance of decile portfolios based on predicted MDD with annual rebalancing. We display in Figure \ref{fig:quantile_noESG} the cumulative returns of the equally-weighted portfolios and report in Table \ref{tab:table_perf_noESG_mdd} some statistics of these portfolios with XGBoost as a prediction model. The results are consistent with other models\footnote{Market-value-weighted portfolios exhibit the same pattern as equally weighted portfolios}. Sharpe and Calmar ratios are mostly decreasing as a function of the quintiles which is desirable for a portfolio manager.
\begin{table}[H]
\centering
	\caption{Portfolio performance metrics for equally-weighted portfolios for XGBoost  (MDD)}
	\caption*{The average of daily observed annualized return (RET), volatility (VOL), maximum drawdown (MDD), and Sharpe and Calmar ratios for the period 2001-2019 for decile portfolios and the market (mkt) for equally-weighted portfolios (XGBoost).}
\begin{tabular}{l|rrrrr}
\toprule
{} &    RET &    VOL &    MDD &    Sharpe Ratio &    Calmar Ratio \\
\midrule
mkt       &  11.04 &  20.16 &  18.43 &  0.49 &  1.04 \\
\midrule
decile 1  &  11.21 &  14.74 &  11.81 &  0.68 &  1.21 \\
decile 2  &  11.36 &  16.88 &  13.30 &  0.65 &  1.34 \\
decile 3  &  11.16 &  17.86 &  14.36 &  0.58 &  1.19 \\
decile 4  &  11.72 &  18.90 &  15.92 &  0.59 &  1.22 \\
decile 5  &  11.27 &  19.76 &  17.18 &  0.55 &  1.17 \\
decile 6  &  12.27 &  20.76 &  18.54 &  0.57 &  1.22 \\
decile 7  &  11.09 &  22.28 &  20.45 &  0.46 &  1.00 \\
decile 8  &  11.61 &  23.61 &  22.81 &  0.46 &  1.00 \\
decile 9  &  10.18 &  26.13 &  26.36 &  0.34 &  0.78 \\
decile 10 &  10.54 &  28.96 &  31.44 &  0.27 &  0.75 \\
\bottomrule
\end{tabular}
 
\label{tab:table_perf_noESG_mdd}
\end{table}
\newpage
\noindent Figure \ref{fig:quantile_noESG} shows how the low MDD decile (decile 1) outperformed other deciles and that it has a lower risk. This result is in line with feature importance, which suggests that volatility is the main factor and that a low MDD portfolio is more likely to be a low volatility portfolio, and vice versa. When we look at Table \ref{tab:table_perf_noESG_mdd}, we can see that the returns of equally-weighted portfolios were higher for lower MDD deciles and that the volatility, maximum drawdown, and Sharpe and Calmar ratios were better for low MDD deciles\footnote{Qualitatively similar results are observed for market-cap-weighted portfolios.}.

\subsection{The impact of ESG scores}
\noindent
Having investigated the ability of non-ESG firm characteristics to predict the ranking of returns and maximum drawdown on a large test set, we repeat the analysis, adding ESG to the mix. Our goal is to determine whether indicators of a company's sustainability improve our ability to predict returns or maximum drawdown.  Since the ESG data set begins only in 2009 for OWL Analytics and 2007 for TVL, we are forced to start our training set later and extend it beyond the financial crisis, providing the models with enough observations with ESG scores. As such, we drop observations without any ESG score and update the models each year to account for more recent data. Using the most recent trained model, we predict one year of out-of-sample data. We also keep variables that improved the predictive models from the previous section which sums up to 72 covariates for log excess returns, and 67 variables for the maximum drawdown. We made sure this validation process uses data before our out-of-sample to avoid look-ahead bias.\\
\\
We compare the results based on the out-of-sample MSE and Kendall correlation as before. For this case, we compare the base case, which uses only non-ESG characteristics, to cases where we add ESG variables. We separate the addition of 4 ESG variables from OWL into two subcases\footnote{This is because the ESG score is constructed as a weighted average of the three pillars E, S, and G scores, it is .} and consider 3 TVL scores. Notations and definitions of these cases are as follows:
\begin{itemize}
	\item Base: Top firm characteristics except for ESG scores.
	\item E/S/G: E, S, and G scores from OWL Analytics.
	\item ESG: Single OWL ESG score.
	\item TVL: volume, insight, and momentum from TruValue Labs data.
\end{itemize}
Using the firm characteristics of the base case and adding different combinations of OWL ESG variables and TVL scores results in six cases of interest. We train these six cases with all the models again but retrain after each year of prediction to increase the number of observations and use more recent market conditions.
\newpage
\subsubsection{The case of returns}
\subsubsection*{Performance}
\begin{figure}[H]
\caption{The evolution of Kendall correlation for the base case (left) and the difference to the base case (right) in time for XGBoost between 2015-01 and 2019-12 (logER).}
\centering
\includegraphics[scale=0.4]{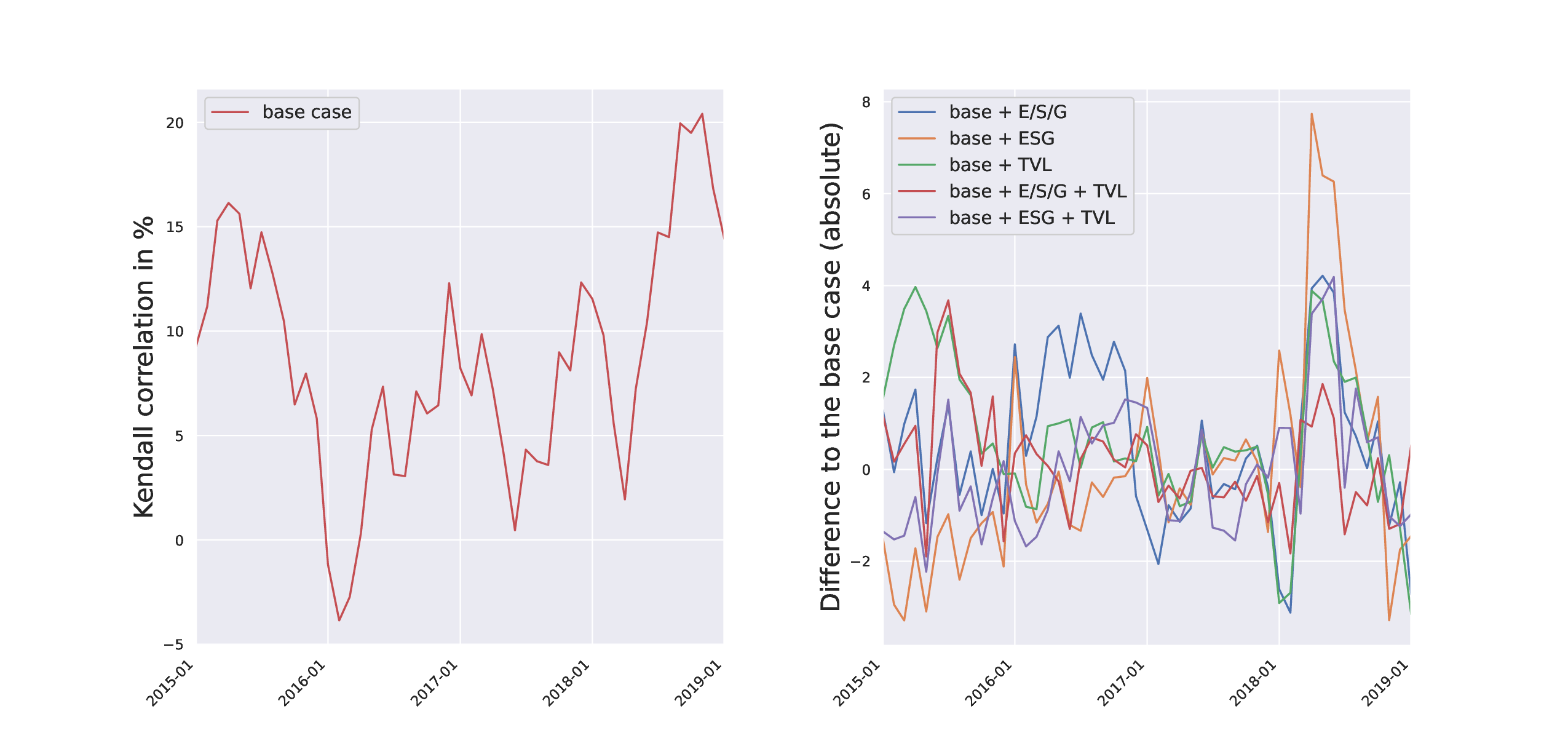}
\label{fig:kendall_time_xgb_ret}
\end{figure}

Table \ref{fig:kendall_tau_ESG} reports the MSE and Kendall correlation for the six different combinations. As indicated above, these results are based on an out-of-sample period between 2015-01 and 2019-12. Without detailing the results, which are reported in the table, we point out the three following insights:
\begin{itemize}
    \item Due to the short training period (at least at the beginning of the period), machine learning methods such as random forest, XGBoost, or MLP underperform linear models at times.
    \item Linear models don't seem to account for ESG variables as seen by the unchanged performance measures.
    \item XGBoost performed slightly better than linear models and MLP, but this outperformance is very small ($\sim$1\%). 
    \item ESG variables have a mixed effect on non-linear models (XGBoost and MLP), sometimes improving the prediction and sometimes causing it to deteriorate.
\end{itemize}
\begin{table}[H]
	\caption{Out-of-sample mean squared error and Kendall correlation (logER)}
	\caption*{The table and barplots report the overall Kendall correlation for the eight different cases, and cover non-linear models; i.e., random forest (RF), XGBoost, and multi-layer perceptron (MLP). We report in Figure (b) some of the results in Table (a), i.e., the base case with the 72 non-ESG firm characteristics (base), and different combinations with ESG variables.}
    \begin{subfigure}{1\textwidth}
    \centering
		\caption{Table of out-of-sample overall MSE}
		\begin{tabular}{llllllllll}
\toprule
{} &   OLS & Lasso & Ridge &  ENet &   PCR &   PLS &    RF & XGBoost &   MLP \\
\midrule
base               & 0.195 & 0.195 & 0.194 & 0.195 & 0.194 & 0.195 & 0.205 &   0.353 & 0.200 \\
base + E/S/G       & 0.195 & 0.195 & 0.195 & 0.195 & 0.194 & 0.195 & 0.202 &   0.316 & 0.200 \\
base + ESG         & 0.195 & 0.195 & 0.195 & 0.195 & 0.194 & 0.195 & 0.202 &   0.336 & 0.197 \\
base + TVL         & 0.195 & 0.195 & 0.194 & 0.195 & 0.195 & 0.195 & 0.206 &   0.337 & 0.200 \\
base + E/S/G + TVL & 0.195 & 0.195 & 0.194 & 0.195 & 0.194 & 0.195 & 0.202 &   0.309 & 0.198 \\
base + ESG + TVL   & 0.195 & 0.195 & 0.194 & 0.195 & 0.194 & 0.195 & 0.202 &   0.325 & 0.198 \\
\bottomrule
\end{tabular}

    \end{subfigure}
    \begin{subfigure}{1\textwidth}
    \centering
		\caption{Table of out-of-sample overall Kendall correlation}
		\begin{tabular}{llllllllll}
\toprule
{} &  OLS & Lasso & Ridge & ENet &   PCR &  PLS &   RF & XGBoost &  MLP \\
\midrule
base         & 7.53 &  9.58 &  8.37 & 7.86 & 10.40 & 7.98 & 9.58 &    8.69 & 7.19 \\
base + E/S/G       & 7.37 &  9.58 &  8.23 & 7.74 & 10.35 & 7.78 & 9.62 &    9.23 & 6.13 \\
base + ESG          & 7.59 &  9.58 &  8.43 & 7.92 & 10.40 & 8.02 & 9.63 &    8.67 & 7.24 \\
base + TVL          & 7.61 &  9.58 &  8.41 & 7.92 & 10.26 & 8.04 & 9.56 &    9.39 & 5.91 \\
base + E/S/G + TVL & 7.44 &  9.58 &  8.26 & 7.79 & 10.41 & 7.84 & 9.63 &  8.85 & 8.11 \\
base + ESG + TVL   & 7.65 &  9.58 &  8.46 & 7.96 & 10.43 & 8.07 & 9.65 &    8.67 & 6.89 \\
\bottomrule
\end{tabular}

    \end{subfigure}   
\label{fig:kendall_tau_ESG}
\end{table}
\newpage
\noindent To explore any potential improvement over time of ESG data, we report the time series of Kendall correlation and compare the base model with the enhanced ESG models between 2015-01 and 2019-12. The time series of the performance metric for XGBoost, which seemed to be the model that capture ESG data the most, are shown in Figure \ref{fig:kendall_time_xgb_ret}.
The change in the performance when including ESG variables varies around zero and it is therefore difficult to associate ESG variables with an improvement in the prediction.
\color{black}
\subsubsection*{Feature importance}
\begin{figure}[h]
\caption{Variable importance for ''base + E/S/G + TVL'' (logER)}
\caption*{The figure reports variable importance defined by the improvement of Kendall correlation when a given variable deviates from the mean.  The results are reported for the nine models for case ``base + E/S/G + TVL'' over the out-of-sample period between 2015-01 and 2019-12. 
}
\centering
\includegraphics[height=15cm, width=12cm]{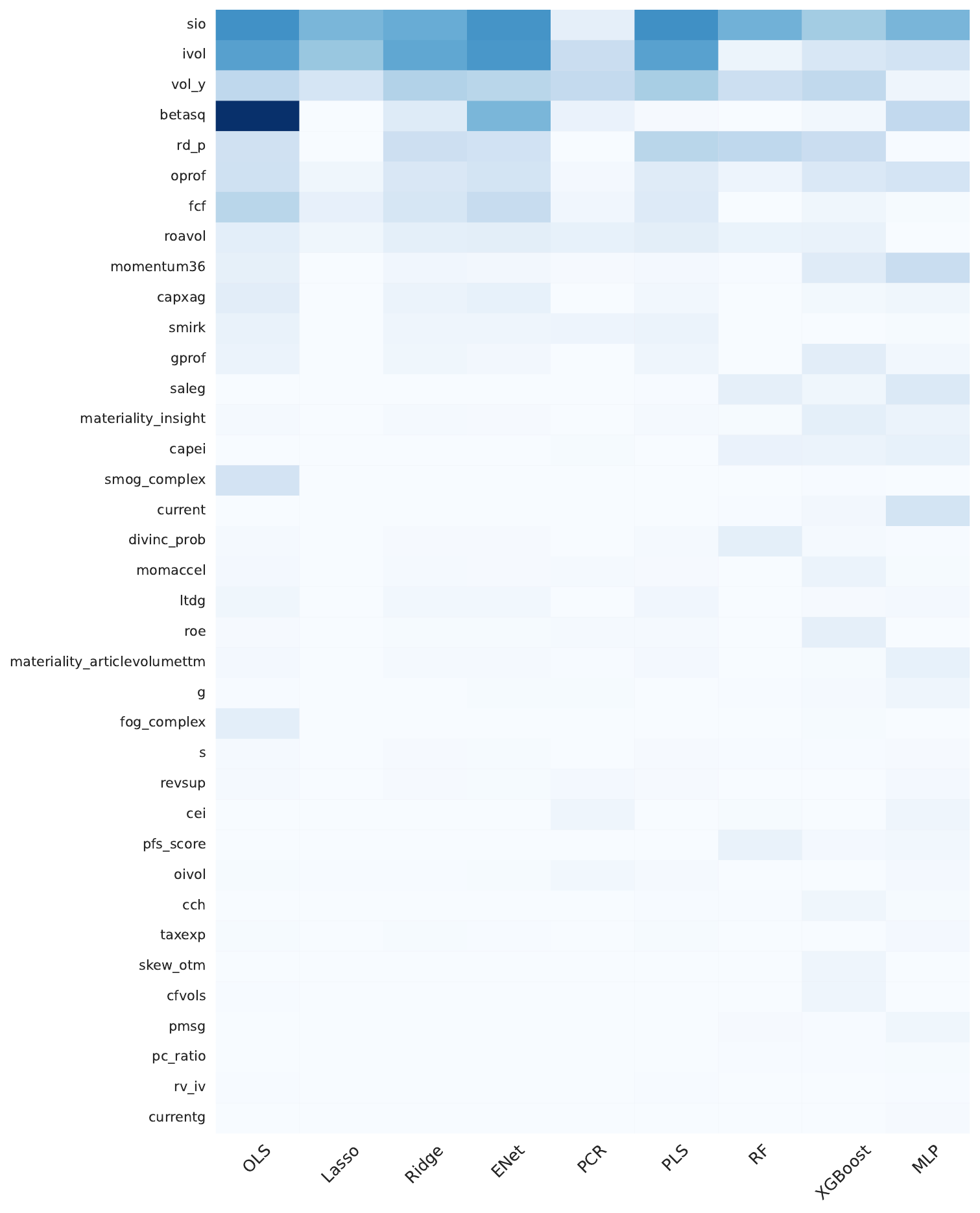}
\label{fig:feature_selection_case0_ret}
\end{figure}

We repeat the feature importance analysis for ``base + E/S/G + TVL'' for all nine models. 
Figure \ref{fig:feature_selection_case0_ret} reports the ranked variables by the average importance score across the nine models. We find again the same top 3 variables, namely, idiosyncratic volatility, realized volatility, and short interest. Among ESG scores, we find TVL's insight score in the top 20, mainly captured by XGBoost. The G score is ranked 23rd among the 72 variables, and the S score is 25th. 
\newpage
\subsubsection*{Portfolio performance}
\begin{figure}[H]
\caption{Cumulative returns of equally-weighted quintile portfolios over time b/w 2015 and 2019 (logER).}
\caption*{Equally-weighted decile portfolios using XGBoost predicted logER deciles from lowest to highest. Portfolios are rebalanced yearly and held for one year. Solid lines (resp. dashed lines) represent the ``base'' case (resp. ``base + E/S/G + TVL''). The market portfolio (mkt) is an equally-weighted portfolio with a yearly rebalancing frequency.}
\centering
\includegraphics[scale=0.4]{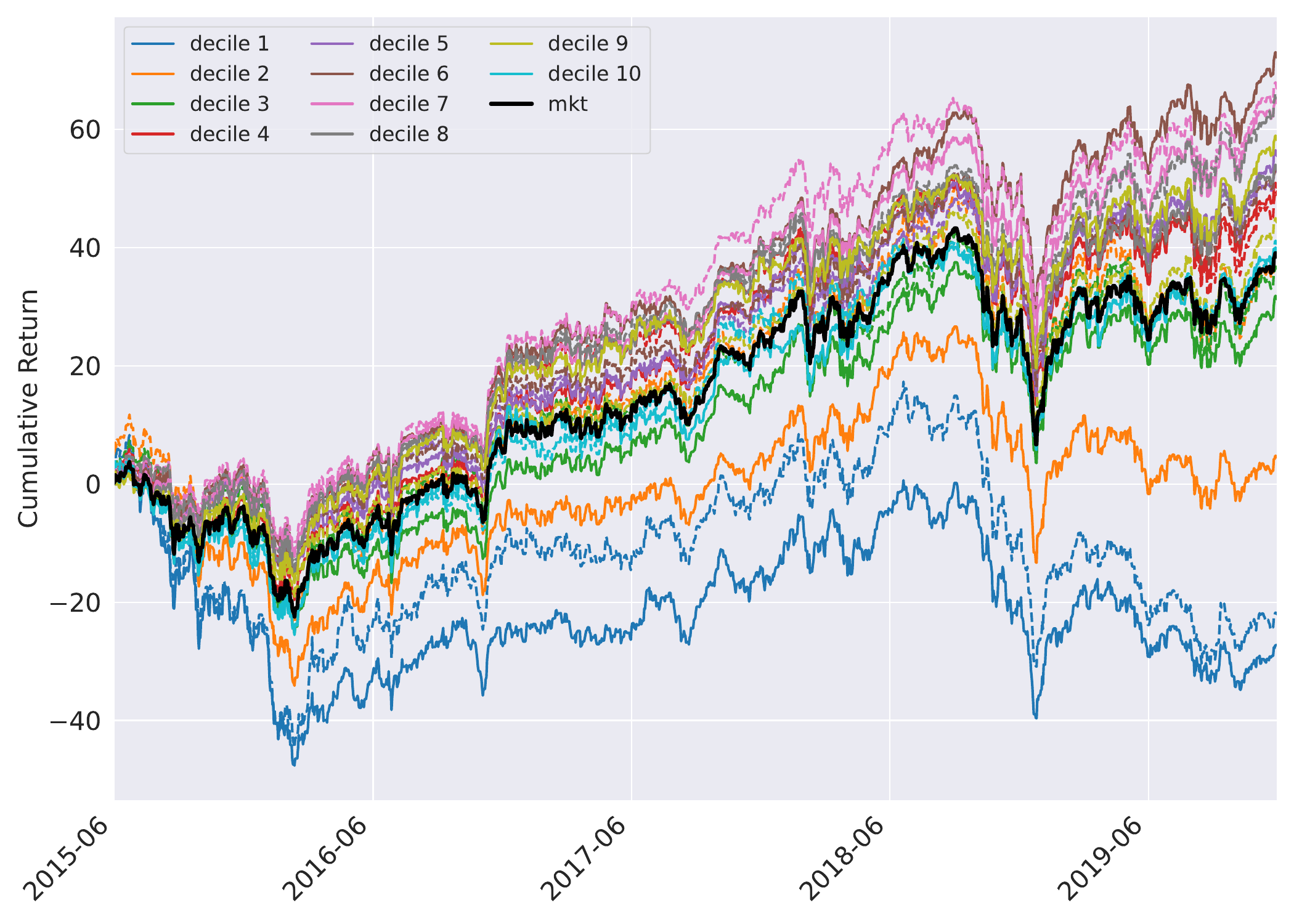}
\label{fig:quantiles_ESG_ret}
\end{figure}

We investigate the performance of decile portfolios based on predicted logER, when ESG variables are included in the set of predictors. We display in Figure \ref{fig:quantiles_ESG_ret} the  cumulative return of equally-weighted decile portfolios, and we report in Table \ref{tab:esg_stats_ew_xgboost_ret} some statistics for those portfolios. \\
\noindent The inclusion of ESG variables made very little difference in the  performance of decile portfolios (see Figure \ref{fig:quantiles_ESG_ret}).  This is consistent with the results shown above that  ESG variables did not significantly improve the prediction of the logER ranking. Table \ref{tab:esg_stats_ew_xgboost_ret}, which reports the average returns, volatility, maximum drawdown, and Sharpe and Calmar ratios, gives an extensive comparison of different cases for one illustrative model. We see a slight difference in decile portfolios. Moreover, including the ESG score seems to increase the gap between the performance of decile 1 and decile 10 which supports the finding that ``base + ESG'' performs better than ``base'' for the XGBoost model.
\begin{table}[H]
	\caption{ESG performance of quintile portfolios at the time of rebalancing (logER)}
	\caption*{Average return, volatility, maximum drawdown, Sharpe and Calmar ratios for decile portfolios between 2015-06 and 2019-12.}
\resizebox{0.7\width}{!}{\begin{tabular}{l|l|rrrrrrrrrrr}
\toprule
   &                  &  decile 1 &  decile 2 &  decile 3 &  decile 4 &  decile 5 &  decile 6 &  decile 7 &  decile 8 &  decile 9 &  decile 10 &  1st - 10th \\
Statistics & Case &           &           &           &           &           &           &           &           &           &            &             \\
\midrule
RET & base &      1.37 &      6.51 &     10.21 &     12.70 &     12.45 &     14.75 &     13.56 &     12.15 &     13.34 &      11.41 &       10.04 \\
   & base + E/S/G &      1.19 &      8.16 &     10.16 &     12.20 &     12.92 &     14.57 &     12.05 &     12.24 &     13.07 &      11.90 &       10.71 \\
   & base + ESG &      0.79 &      7.83 &      9.17 &     13.08 &     13.16 &     13.78 &     13.93 &     11.72 &     12.61 &      12.40 &       11.61 \\
   & base + TVL &     -0.60 &      7.88 &     12.66 &     11.57 &     12.63 &     12.90 &     14.15 &     12.74 &     11.99 &      12.55 &       13.15 \\
   & base + E/S/G + TVL &      0.02 &      8.11 &     10.43 &     10.80 &     14.43 &     13.53 &     14.52 &     12.59 &     11.03 &      13.01 &       12.99 \\
   & base + ESG + TVL &     -0.73 &      7.13 &     11.93 &     11.70 &     14.44 &     12.48 &     13.72 &     11.68 &     11.64 &      14.36 &       15.09 \\
\midrule
VOL & base &     23.60 &     18.50 &     16.07 &     15.24 &     14.21 &     13.89 &     13.51 &     13.71 &     13.90 &      15.48 &       -8.12 \\
   & base + E/S/G &     23.45 &     18.21 &     15.80 &     14.85 &     14.43 &     13.83 &     13.69 &     13.91 &     14.21 &      15.42 &       -8.03 \\
   & base + ESG &     23.42 &     18.00 &     16.20 &     14.76 &     14.06 &     13.95 &     13.69 &     13.68 &     14.29 &      15.84 &       -7.58 \\
   & base + TVL &     23.84 &     18.00 &     15.78 &     14.77 &     14.03 &     13.85 &     13.68 &     14.34 &     13.99 &      15.65 &       -8.19 \\
   & base + E/S/G + TVL &     23.27 &     18.27 &     16.16 &     14.82 &     14.03 &     13.91 &     13.72 &     13.89 &     14.03 &      15.69 &       -7.58 \\
   & base + ESG + TVL &     23.83 &     18.20 &     16.15 &     15.46 &     14.29 &     13.82 &     13.63 &     13.52 &     13.61 &      15.43 &       -8.40 \\
\midrule
MDD & base &     25.49 &     19.69 &     15.46 &     14.45 &     13.65 &     12.38 &     12.22 &     12.36 &     12.42 &      14.93 &      -10.56 \\
   & base + E/S/G &     25.11 &     18.75 &     15.96 &     14.14 &     13.89 &     12.34 &     12.58 &     12.46 &     12.74 &      14.96 &      -10.15 \\
   & base + ESG &     25.15 &     18.89 &     15.95 &     14.14 &     13.51 &     12.63 &     11.98 &     12.36 &     12.98 &      15.16 &       -9.99 \\
   & base + TVL &     25.54 &     18.64 &     15.34 &     14.60 &     13.06 &     13.06 &     12.23 &     12.82 &     12.53 &      14.95 &      -10.59 \\
   & base + E/S/G + TVL &     24.72 &     19.54 &     15.57 &     14.45 &     12.77 &     12.60 &     12.36 &     12.55 &     12.99 &      14.95 &       -9.77 \\
   & base + ESG + TVL &     25.59 &     19.16 &     15.99 &     14.82 &     13.16 &     12.78 &     12.17 &     12.21 &     12.59 &      14.33 &      -11.26 \\
\midrule
Sharpe Ratio & base &      0.06 &      0.31 &      0.56 &      0.75 &      0.81 &      0.95 &      0.90 &      0.75 &      0.84 &       0.64 &        0.58 \\
   & base + E/S/G &      0.04 &      0.42 &      0.56 &      0.72 &      0.81 &      0.94 &      0.76 &      0.77 &      0.83 &       0.68 &        0.64 \\
   & base + ESG &      0.02 &      0.41 &      0.49 &      0.80 &      0.86 &      0.88 &      0.91 &      0.73 &      0.77 &       0.71 &        0.69 \\
   & base + TVL &     -0.04 &      0.41 &      0.71 &      0.68 &      0.80 &      0.83 &      0.91 &      0.79 &      0.76 &       0.71 &        0.75 \\
   & base + E/S/G + TVL &     -0.02 &      0.42 &      0.56 &      0.63 &      0.92 &      0.85 &      0.97 &      0.78 &      0.68 &       0.75 &        0.77 \\
   & base + ESG + TVL &     -0.05 &      0.35 &      0.66 &      0.65 &      0.92 &      0.81 &      0.89 &      0.77 &      0.75 &       0.85 &        0.90 \\
\midrule
Calmar Ratio & base &      0.37 &      0.65 &      1.00 &      1.39 &      1.41 &      1.57 &      1.60 &      1.45 &      1.45 &       1.10 &        0.73 \\
   & base + E/S/G &      0.32 &      0.85 &      0.99 &      1.25 &      1.30 &      1.74 &      1.29 &      1.47 &      1.52 &       1.14 &        0.82 \\
   & base + ESG &      0.30 &      0.80 &      0.95 &      1.44 &      1.45 &      1.58 &      1.53 &      1.28 &      1.36 &       1.22 &        0.92 \\
   & base + TVL &      0.22 &      0.91 &      1.34 &      1.16 &      1.34 &      1.43 &      1.58 &      1.44 &      1.32 &       1.20 &        0.98 \\
   & base + E/S/G + TVL &      0.28 &      0.89 &      0.99 &      1.09 &      1.65 &      1.43 &      1.65 &      1.41 &      1.23 &       1.30 &        1.02 \\
   & base + ESG + TVL &      0.23 &      0.75 &      1.08 &      1.10 &      1.71 &      1.52 &      1.63 &      1.41 &      1.40 &       1.34 &        1.11 \\
\bottomrule
\end{tabular}
}
\label{tab:esg_stats_ew_xgboost_ret}
\end{table}

\subsubsection{The case of maximum drawdown}
\subsubsection*{Performance}
\begin{figure}[h]
\caption{The evolution of Kendall correlation for the base case (left) and the difference to the base case (right) in time for MLP between 2015-01 and 2019-12 (MDD).}
\centering
\includegraphics[scale=0.35]{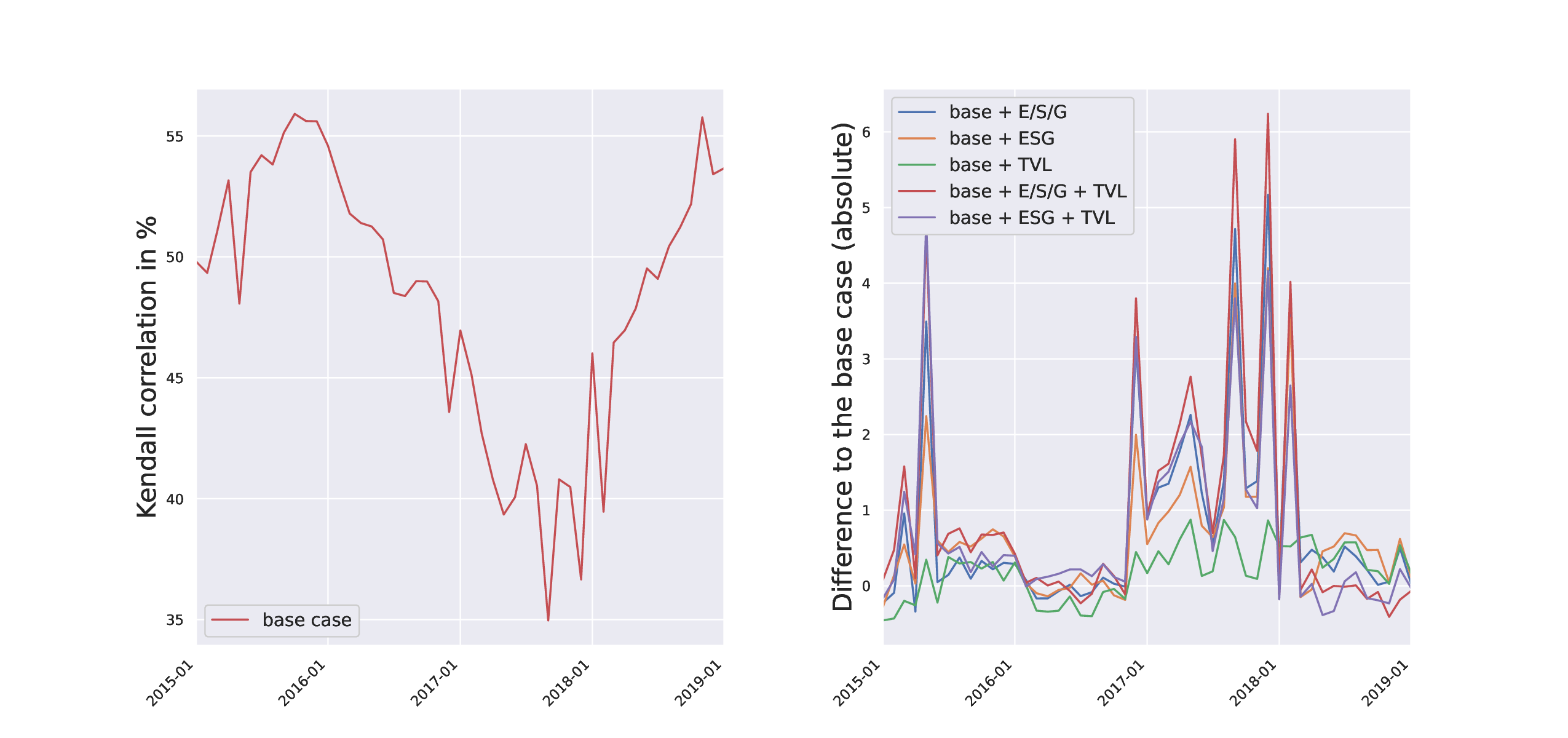}
\label{fig:kendall_time_xgb}
\end{figure}

Using the same out-of-sample period between 2015-01 and 2019-12, we report in Table \ref{fig:kendall_tau_ESG_mdd} the MSE and Kendall correlation for the six different combinations and nine different models. We point out the following insights:
\begin{itemize}
    \item Results for the new test sample period (2015-2019) were better than for the previous test period (2001 to 2019). This is not surprising because of the absence of the 9 crisis and the shorter time period.
    \item XGBoost performed slightly better than other models when looking at the correlation but this outperformance is very small ($\sim$2-4\% relative) but underperformed in terms of MSE. Again, this result is not surprising given that the training set is relatively small. 
    \item ESG variables marginally improved the results for non-linear models (<2\% relative).
    One explanation for that could be that the information for ranking the stocks by their MDD is already explained by the other variables, especially volatility-based characteristics, or that they simply have a very low predictive power of MDD.
\end{itemize}
\newpage
\noindent Although these results show an absence of improvement of maximum drawdown prediction using ESG variables, we emphasize that this result is bound to the set of characteristics used along with ESG data as well as the universe of stocks used. It is important to also stress that these variables are quite recent and missing for multiple data points. 
\begin{table}[H]
	\caption{Out-of-sample mean squared error and Kendall correlation (MDD)}
	\caption*{The table and barplots report the overall Kendall correlation for the eight different cases, and cover non-linear models; i.e., random forest (RF), XGBoost, and multi-layer perceptron (MLP). We report in the Figure (b) some of the results in Table (a), i.e., the base case with the 98 non-ESG firm characteristics (base), the base case along with E, S, and G OWL scores (base + OWL E/S/G),  the base with TVL four scores (base + TVL scores), and the base with E, S, G OWL scores and TVL scores (base + E/S/G + TVL scores).}
    \begin{subfigure}{1\textwidth}
    \centering
		\caption{Table of out-of-sample MSE}
		\begin{tabular}{llllllllll}
\toprule
{} &   OLS & Lasso & Ridge &  ENet &   PCR &   PLS &    RF & XGBoost &   MLP \\
\midrule
base               & 0.020 & 0.020 & 0.020 & 0.020 & 0.020 & 0.020 & 0.025 &   0.024 & 0.021 \\
base + E/S/G       & 0.020 & 0.020 & 0.020 & 0.020 & 0.020 & 0.020 & 0.025 &   0.022 & 0.021 \\
base + ESG         & 0.020 & 0.020 & 0.020 & 0.020 & 0.020 & 0.020 & 0.025 &   0.023 & 0.021 \\
base + TVL         & 0.020 & 0.020 & 0.020 & 0.020 & 0.020 & 0.020 & 0.025 &   0.024 & 0.021 \\
base + E/S/G + TVL & 0.020 & 0.020 & 0.020 & 0.020 & 0.020 & 0.020 & 0.025 &   0.022 & 0.020 \\
base + ESG + TVL   & 0.020 & 0.020 & 0.020 & 0.020 & 0.020 & 0.020 & 0.025 &   0.023 & 0.021 \\
\bottomrule
\end{tabular}

    \end{subfigure}
    \begin{subfigure}{1\textwidth}
    \centering
		\caption{Table of out-of-sample average Kendall correlation}
		\begin{tabular}{llllllllll}
\toprule
{} &    OLS &  Lasso &  Ridge &   ENet &    PCR &    PLS &     RF & XGBoost &    MLP \\
\midrule
base         & 47.91 & 48.07 & 48.16 & 48.00 & 47.19 & 47.95 & 45.95 &  48.31 & 46.93 \\
base + E/S/G       & 47.96 & 48.12 & 48.20 & 48.04 & 47.26 & 47.97 & 46.176 &  49.10 & 46.81 \\
base + ESG          & 47.96 & 48.13 & 48.22 & 48.05 & 47.26 & 48.01 & 46.17 &  49.02 & 46.80 \\
base + TVL          & 48.01 & 48.18 & 48.26 & 48.10 & 47.43 & 48.04 & 45.944 &  48.51 & 46.82 \\
base + E/S/G + TVL & 48.03 & 48.20 & 48.28 & 48.12 & 47.51 & 48.04 & 46.16 &  49.29 & 46.72 \\
base + ESG + TVL   & 48.04 & 48.20 & 48.29 & 48.13 & 47.47 & 48.07 & 46.15 &  49.07 & 47.14 \\
\bottomrule
\end{tabular}

    \end{subfigure}   
\label{fig:kendall_tau_ESG_mdd}
\end{table}
\noindent We explore how the models performed on different dates through the time series of cross-sectional Kendall correlation between 2015-01 and 2019-12 (see Figure \ref{fig:kendall_time_xgb}).  
We find that including ESG variables led to a significant improvement in some dates (almost 10\% relative). The difference in performance between the base case and ``base + E/S/G + TVL'' is mostly positive throughout the period. 
\color{black}
\subsubsection*{Feature importance}
\begin{figure}[H]
\caption{Variable importance for MLP and 4 combination of cases (MDD)}

\caption*{
The figure reports variable importance defined by the improvement of Kendall correlation when a given variable deviates from the mean.  The results are reported for the nine models for case ``base + E/S/G + TVL'' over the out-of-sample period between 2015-01 and 2019-12.
}
\centering
\includegraphics[height=15cm, width=12cm]{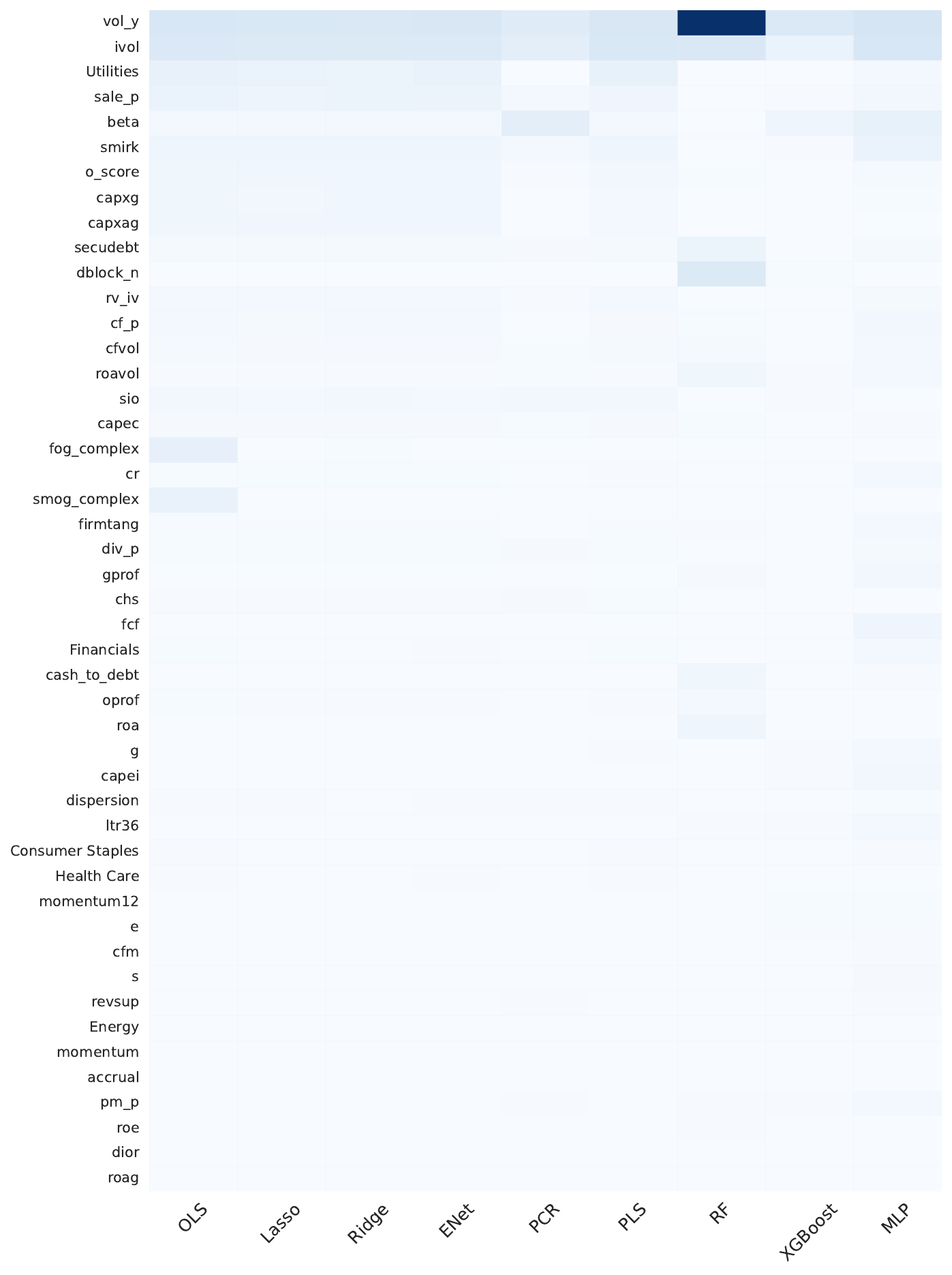}
\label{fig:feature_selection_case0_mdd}
\end{figure}

We compute feature importance for the case ``base + E/S/G + TVL'' in order to assess the predictive power of ESG variables for all nine models. Figure \ref{fig:feature_selection_case0_mdd} reports the variables ranked by the average importance scores. The leading dominant variables across the models were still the same, i.e., one-year volatility (vol\_y), idiosyncratic volatility (ivol) and beta. The G score comes 30th out of a total of 74, 47 of which don't decrease the performance.
\subsubsection*{Portfolio performance}
\begin{figure}[ht]
\caption{Cumulative returns of equally-weighted decile portfolios over time b/w 2015 and 2019 (MDD)}
\caption*{Equally-weighted decile portfolios using XGBoost predicted MDD deciles from lowest to highest. Portfolios are rebalanced yearly and held for one year. Solid lines (resp. dashed lines) represent the ``base'' case (resp. ``base + E/S/G + TVL''). The market portfolio (mkt) is an equally-weighted portfolio with a yearly rebalancing frequency.
}
\centering
\includegraphics[scale=0.4]{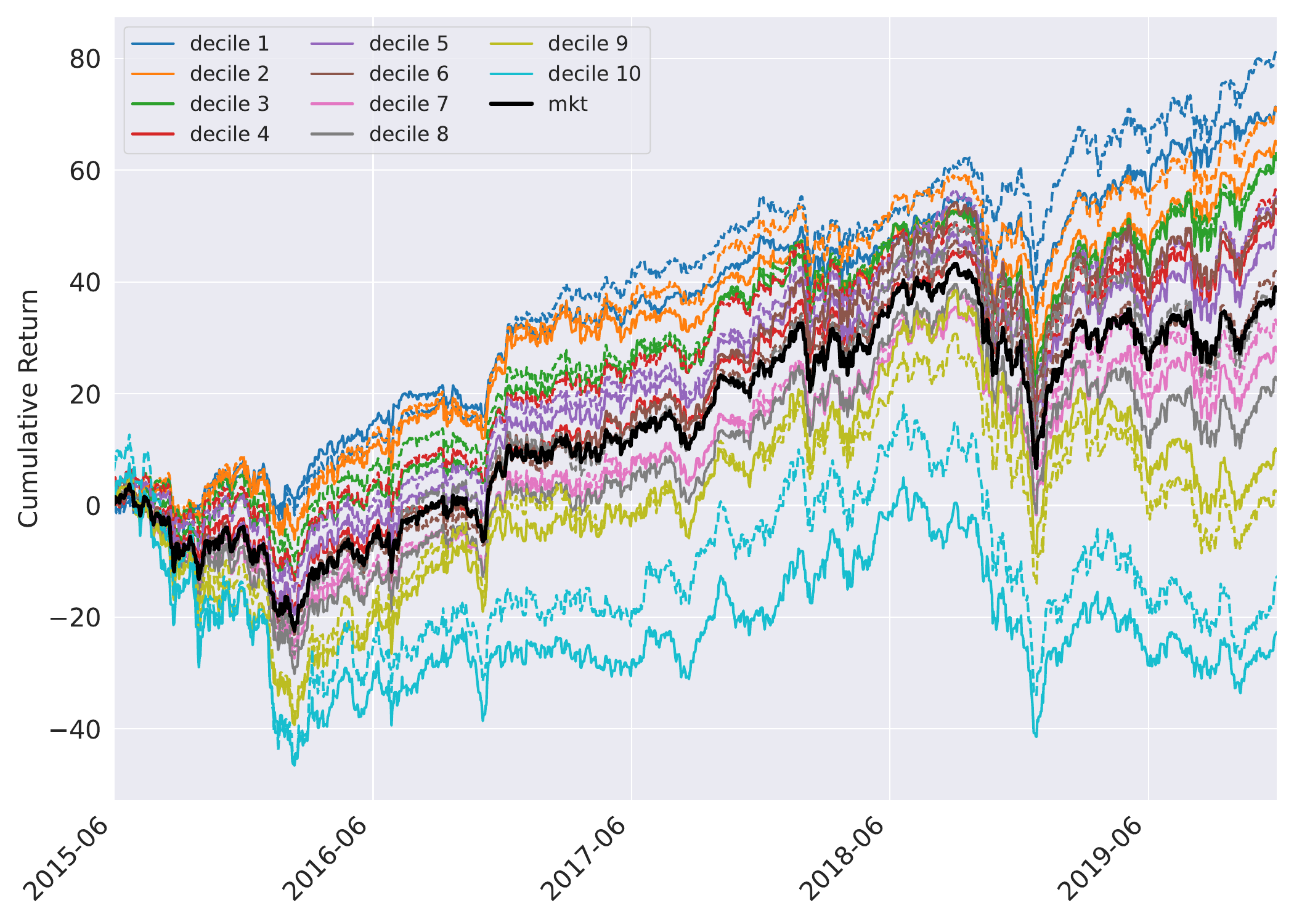}
\label{fig:quantiles_ESG_mdd}
\end{figure}

We finally investigate the performance of decile portfolios based on predicted MDD, where ESG variables are included in the prediction. 
We display in Figure \ref{fig:quantiles_ESG_mdd} the  cumulative return to equally-weighted quintile portfolio, and we report in Table \ref{tab:esg_stats_ew_xgboost_mdd_esg}, some statistics for both equally-weighted and market-cap-weighted quintile portfolios. \\
\newline
\noindent The inclusion of ESG variables made visually no difference in the  performance of decile portfolios (see Figure \ref{fig:quantiles_ESG_mdd}).  This is consistent with the results shown above that  ESG variables did not significantly improve the prediction of the MDD ranking. This is also consistent with the results in Table \ref{tab:esg_stats_ew_xgboost_mdd_esg} which report the average returns, volatility, maximum drawdown, and Sharpe and Calmar ratios. There is very little change in the decile portfolio performance when including ESG variables.
\\
\begin{table}[H]
	\caption{ESG performance of quintile portfolios at the time of rebalancing (MDD)}
	\caption*{Average return, volatility, maximum drawdown, Sharpe and Calmar ratios for decile portfolios between 2015-06 and 2019-12.}
\resizebox{0.7\width}{!}{\begin{tabular}{l|l|rrrrrrrrrrr}
\toprule
   &                  &  decile 1 &  decile 2 &  decile 3 &  decile 4 &  decile 5 &  decile 6 &  decile 7 &  decile 8 &  decile 9 &  decile 10 &  10th - 1st \\
Statistics & Case &           &           &           &           &           &           &           &           &           &            &             \\
\midrule
RET & base &     15.33 &     12.54 &      9.04 &      9.76 &      9.18 &      9.74 &      8.38 &      8.13 &      5.63 &      -3.01 &       18.34 \\
   & base + E/S/G &     14.70 &     12.01 &     10.18 &      8.46 &     10.54 &      8.62 &      8.81 &      9.48 &      3.34 &      -1.34 &       16.04 \\
   & base + ESG &     13.84 &     12.03 &     10.75 &      8.82 &     10.89 &      8.52 &      9.45 &      8.95 &      3.60 &      -2.02 &       15.86 \\
   & base + TVL &     15.33 &     12.14 &     10.01 &      8.99 &      9.06 &     10.02 &      8.28 &      8.61 &      4.48 &      -2.17 &       17.50 \\
   & base + E/S/G + TVL &     14.97 &     11.56 &     10.01 &      9.08 &     10.32 &      8.04 &      9.10 &      9.02 &      4.52 &      -1.86 &       16.83 \\
   & base + ESG + TVL &     14.74 &     12.09 &      9.53 &      9.62 &      9.76 &      8.70 &      9.00 &      8.84 &      4.41 &      -2.02 &       16.76 \\
\midrule
VOL & base &     11.32 &     12.19 &     13.34 &     14.09 &     14.98 &     15.87 &     16.97 &     18.53 &     21.69 &      28.97 &      -17.65 \\
   & base + E/S/G &     11.21 &     12.36 &     13.29 &     14.03 &     14.96 &     15.83 &     17.00 &     18.74 &     21.76 &      29.01 &      -17.80 \\
   & base + ESG &     11.25 &     12.17 &     13.38 &     14.04 &     15.02 &     15.80 &     17.04 &     18.57 &     21.82 &      29.07 &      -17.82 \\
   & base + TVL &     11.21 &     12.28 &     13.29 &     14.15 &     14.99 &     15.76 &     17.00 &     18.67 &     21.86 &      28.90 &      -17.69 \\
   & base + E/S/G + TVL &     11.21 &     12.33 &     13.38 &     13.97 &     14.99 &     15.77 &     16.89 &     18.83 &     21.75 &      29.05 &      -17.84 \\
   & base + ESG + TVL &     11.26 &     12.32 &     13.44 &     13.81 &     15.01 &     15.85 &     16.84 &     18.71 &     21.89 &      29.09 &      -17.83 \\
\midrule
MDD & base &      8.89 &     10.31 &     12.66 &     13.32 &     14.60 &     15.79 &     17.01 &     17.69 &     23.42 &      30.37 &      -21.48 \\
   & base + E/S/G &      8.53 &     10.93 &     12.33 &     13.33 &     14.47 &     15.64 &     17.52 &     17.79 &     23.76 &      29.87 &      -21.34 \\
   & base + ESG &      8.78 &     10.59 &     12.32 &     13.53 &     14.21 &     15.90 &     17.14 &     17.93 &     23.67 &      30.08 &      -21.30 \\
   & base + TVL &      8.50 &     10.78 &     12.17 &     13.61 &     14.69 &     15.62 &     17.24 &     18.03 &     23.49 &      30.01 &      -21.51 \\
   & base + E/S/G + TVL &      8.63 &     10.68 &     12.68 &     13.30 &     14.55 &     15.62 &     17.17 &     17.99 &     23.46 &      30.12 &      -21.49 \\
   & base + ESG + TVL &      8.62 &     10.81 &     12.52 &     13.06 &     14.70 &     15.79 &     16.93 &     17.87 &     23.58 &      30.16 &      -21.54 \\
\midrule
Sharpe Ratio & base &      1.24 &      0.89 &      0.58 &      0.61 &      0.54 &      0.57 &      0.47 &      0.38 &      0.25 &      -0.08 &        1.32 \\
   & base + E/S/G &      1.17 &      0.85 &      0.66 &      0.52 &      0.65 &      0.48 &      0.52 &      0.47 &      0.13 &      -0.02 &        1.19 \\
   & base + ESG &      1.10 &      0.86 &      0.70 &      0.55 &      0.67 &      0.49 &      0.53 &      0.44 &      0.14 &      -0.04 &        1.14 \\
   & base + TVL &      1.24 &      0.86 &      0.64 &      0.56 &      0.54 &      0.59 &      0.47 &      0.42 &      0.18 &      -0.05 &        1.29 \\
   & base + E/S/G + TVL &      1.20 &      0.80 &      0.66 &      0.56 &      0.65 &      0.44 &      0.53 &      0.45 &      0.18 &      -0.04 &        1.24 \\
   & base + ESG + TVL &      1.18 &      0.85 &      0.60 &      0.62 &      0.60 &      0.49 &      0.52 &      0.44 &      0.18 &      -0.04 &        1.22 \\
\midrule
Calmar Ratio & base &      2.33 &      1.71 &      1.18 &      1.23 &      1.12 &      1.14 &      0.99 &      0.78 &      0.64 &       0.23 &        2.10 \\
   & base + E/S/G &      2.22 &      1.52 &      1.31 &      1.00 &      1.38 &      0.96 &      1.20 &      0.90 &      0.42 &       0.34 &        1.88 \\
   & base + ESG &      2.03 &      1.53 &      1.39 &      1.05 &      1.42 &      0.96 &      1.14 &      0.87 &      0.44 &       0.31 &        1.72 \\
   & base + TVL &      2.32 &      1.57 &      1.28 &      1.10 &      1.16 &      1.18 &      1.03 &      0.82 &      0.50 &       0.30 &        2.02 \\
   & base + E/S/G + TVL &      2.25 &      1.45 &      1.33 &      1.07 &      1.36 &      0.90 &      1.19 &      0.86 &      0.50 &       0.32 &        1.93 \\
   & base + ESG + TVL &      2.14 &      1.62 &      1.20 &      1.17 &      1.27 &      0.96 &      1.18 &      0.86 &      0.51 &       0.31 &        1.83 \\
\bottomrule
\end{tabular}
}
\label{tab:esg_stats_ew_xgboost_mdd_esg}
\end{table}

\subsubsection{ESG scores contribution to the prediction}
We conclude our analysis by exploring the contribution of ESG variables to the prediction. For this purpose, we use a novel approach to variables' importance, namely the Shapley Additive Explanation (SHAP) (see \cite{LundbergL17}). SHAP relies on Shapley value which is a solution concept in cooperative game theory introduced by \cite{shapley1951notes}. In the context of the model's prediction, explanatory variables are the players in this cooperative game, and the model $f$ plays the role of the coalition whose payoff is the model's prediction.\\
\\
Let us consider a permutation $P$ of the set of indices $\{1,2,\ldots, p\}$ corresponding to an ordering of $p$ explanatory variables included in the model $f$. Denote by $\pi(P, j)$ the set of the indices of the variables that are positioned in $P$ before the $j$-th variable. Note that, if the $j$-th variable is placed as the first, then $\pi(J,j)=\emptyset$. Consider the model's prediction $f(\underbar{x}_{*})$ for a particular instance of interest $\underbar{x}_{*}$. The Shapley value is defined as follows:
\begin{align*}
    \phi(\underbar{x}_{*}, j) = \frac{1}{p!}\sum_{J}\Delta^{j|\pi(J,j)}(\underbar{x}_{*}),
\end{align*}
where the sum is taken over all possible permutations $p!$ (ordering of explanatory variables). The variable-importance measure $\Delta^{j|J}$ for the j-th is the change between the expected prediction, when setting the values of the explanatory variables with indices from the set $J\cup \{j\}$ equal to their values in $\underbar{x}_{*}$, and the expected prediction conditional on setting the values of the explanatory variables with indices from the set $J$ equal to their values in $\underbar{x}_{*}$, i.e.
$\Delta^{j|J}=\mathbb{E}[f(\underline{X})|X^{j_1}=\underline{x}_*^{j_1},\ldots,  X^{j_K}=\underline{x}_*^{j_K},\ldots, X^{j}=\underbar{x}^{l}_{*}] - \mathbb{E}[f(\underline{X})|X^{j_1}=\underline{x}_*^{j_1},\ldots,  X^{j_K}=\underline{x}_*^{j_K}]$.\\
\\
Calculating Shapley value is NP-hard and the time increases exponentially with the number of explanatory variables, however, \cite{LundbergL17} provides an efficient implementation of computations of Shapley values for tree-based models which we will rely on for our application. In our analysis we apply SHAP to the XGBoost model for the reason above, but also because XGBoost is the nonlinear model that seemed to capture the effect of ESG variables through our analysis.

\begin{figure}[H]
\caption{Variables importance}
\caption*{Feature importance for XGBoost and ``base + E/S/G + TVL'' using Shapley value for log excess returns (left) and maximum drawdown (right). Blue bars correspond to a negative impact while red bars correspond to a positive impact. 
}
\centering
\includegraphics[width=8.3cm, height=15cm]{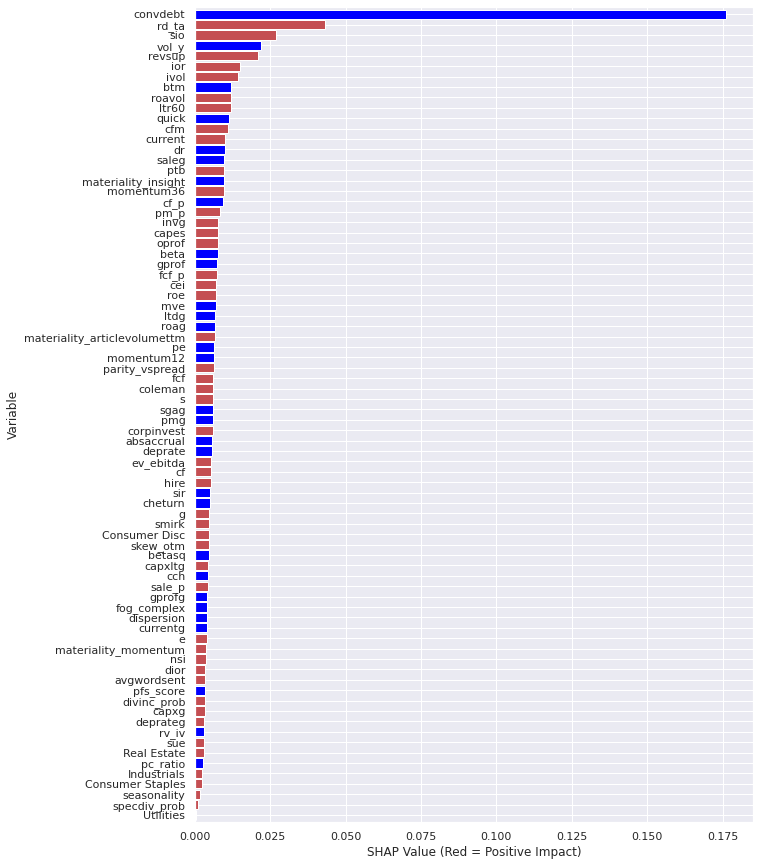}
\includegraphics[width=8.3cm, height=15cm]{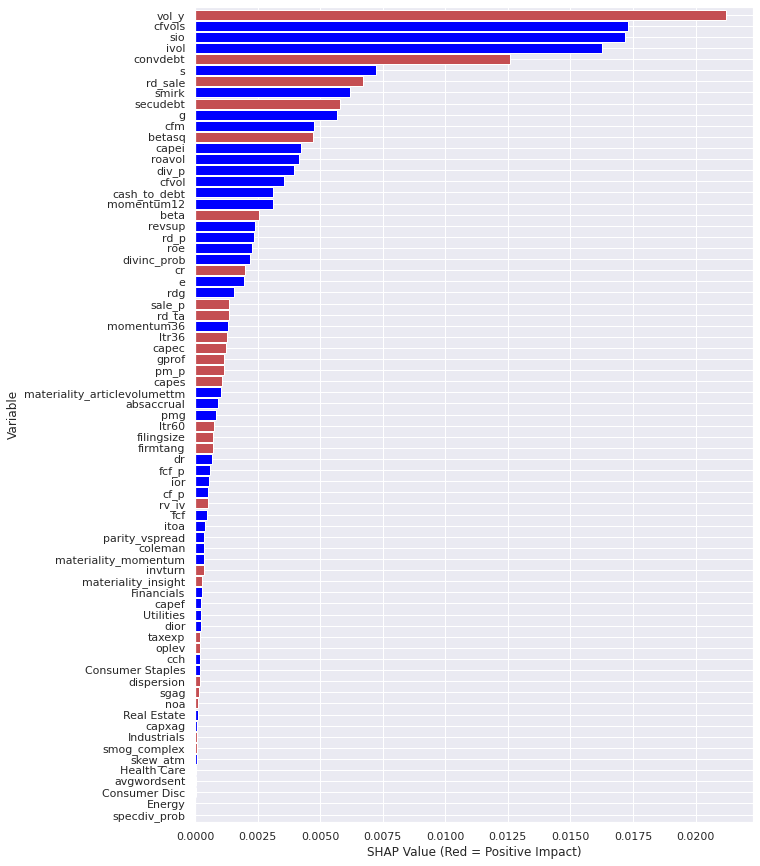}
\label{fig:shap_feature_importance}
\end{figure}
\noindent We can see from Figure  \ref{fig:shap_feature_importance} that TVL's insight score is ranked among the top 20 variables with a positive impact on return, the S score is ranked 40th and the G score is 51st. All these scores are colored in red which suggests a positive impact on returns. For maximum drawdown, the S score is ranked 12th, the G score 15th, and the E score 28th. S and G score have a negative impact on maximum drawdown (negatively correlated with the prediction), however, the E score has a positive correlation with maximum drawdown prediction which suggest that companies with a high E score tend to have a high maximum drawdown. Upon verifying the surprising result, the E score exhibits a small negative correlation with maximum drawdown which explains the feature importance result. As reported by literature, ESG data are divergent among data providers (see \cite{Dimson75}). This complicates such analysis seeking to find the relationship between ESG data and stock risk and performance. 
\section{Conclusion}
This paper explores the cross-section of returns and maximum drawdown in the US equity market. We apply supervised learning methods to forecast the one-year forward returns and maximum drawdown using current information on firm-level characteristics. We find a cross-sectional correlation between the predicted and realized target ranges between 25\% and 60\% with an average of $\sim 45\%$ for the maximum drawdown. For excess returns, the correlation goes between -25\% and 40\% for an average of $\sim 9\%$. This resulted in a good performance of the best decile portfolios and a weak performance of the worse ones in most of the periods of the backtests.\\
\\
In terms of model comparison, the results indicate that non-linear methods slightly outperform linear methods. Some firm characteristics, like idiosyncratic volatility, beta, option volatility, or short interest, can explain the cross-section of both returns and maximum drawdowns. Some factors are more important to returns, like operational profitability or book-to-market. In contrast, others are more relevant to maximum drawdowns, like price-to-sales or profit-margin-to-price. A number of these characteristics are consistently dominant across models, namely one-year volatility and idiosyncratic volatility, beta, and short interest. Finally, the agreement between predicted and realized maximum drawdown persists even in periods of turmoil.\\ 
\\
The addition of environmental, social, and governance scores to the set of predictors failed to improve our models' performance significantly. We identified two variables, the Governance score (g) and the Article Volume (articlevolumettm), captured mainly by the MLP model. These two scores were notably higher in low maximum drawdown/high logER quintile portfolios, consistent with the correlation between the target and those ESG variables. In terms of prediction, ESG variables failed to improve the models in a significant way in our data. This is consistent with other empirical analyses. We might not expect to find significant contributions by ESG variables due to the disparate methodologies by different ESG data providers, and the high correlation between ESG variables and some firm characteristics like size or profitability. To explore the question further, we conducted a Shapley value analysis to measure ESG scores' contribution in cooperative game theory. We found that the contribution of ESG scores, particularly the Social score, was not insignificant on average. We also found that higher scores are associated with higher returns and lower maximum drawdown.\\
\\
Our empirical conclusions about the association between ESG indicators and returns or maximum drawdown must be framed in terms of the data set. Currently, ESG indicators have relatively short data histories, and the growth and evolution of markets and ESG may provide a new perspective in the near term.
\newpage
\appendix

\begin{landscape}
\section{Appendix}
List of firm variables used as predictors from WRDS
\begin{longtable}{llll}
\toprule
Variable Name &                                   Variable Label &   Variable Name &                                     Variable Label \\
\midrule
absaccrual &                               Absolute Accruals  &             ior &                                     Institutional  \\
accrual &                                        Accruals  &            itoa &                              Investment to Assets  \\
accrualpct &                              Accrual Percentage  &            ivol &                                       60-Day IVol  \\
avgwordsent &                             Sentence Complexity  &          liqvol &                           Volatility of Liquidity  \\
beta &                                   12-Month Beta  &            ltdg &                          Total Liabilities Growth  \\
avgwordsent &                             Sentence Complexity  &           ltr36 &                                      36-Month LTR  \\
beta &                                   12-Month Beta  &           ltr60 &                                      60-Month LTR  \\
betasq &                            12-Month Beta Squared &        momaccel &                             Momentum Acceleration  \\
btm &                            Fiscal Book-toMarket  &        momentum &                                  6-Month Momentum  \\
capec &         Cyclically-Adjusted Price to Cash Flows  &      momentum12 &                                 12-Month Momentum  \\
capes & Cyclically-Adjusted Price to Sales &       momentum36 &                                 36-Month Momentum  \\
cashprod &                               Cash Productivity  &             noa &                              Net Operating Assets  \\
cch &                                  Change in Cash  &             nsi &                                Net Stock Issuance  \\
cei &                       Composite Equity Issuance  &         o\_score &                                      Ohlson Score  \\
ceqg &             Growth in Common Stockholder Equity  &           oivol &                       Excess Cash Flow Volatility  \\
cf &                            Cash Flow Efficiency  &           oplev &                    Operating Liabilities Leverage  \\
cf\_p &                             Cash Flows to Price  &           oprof &                           Operating Profitability  \\
cfm &                               Cash Flow Margins  &  parity\_vspread &                 Put-Call Parity Volatility Spread  \\
cfvol &                            Cash Flow Volatility  &   pc\_divergence &                Option-Based Divergence of Opinion  \\
cfvols &                     Cash Flow Margin Volatility  &        pc\_ratio &                                    Put-Call Ratio  \\
cheturn &                                   Cash Turnover  &              pe &                              Price Earnings Ratio  \\
cheturng &                            Cash Turnover Growth  &       pfs\_score &               Piotroski Financial Statement Score  \\
chs &                                    CHS Distress  &            pm\_p &                            Profit Margin to Price  \\
coleman &                       Coleman Readability Index  &             pmg &                     Abnormal Profit Margin Growth  \\
convdebt &                                Convertible Debt  &            pmsg &                               Gross Margin Growth  \\
corpinvest &                           Corporate Investments  &             ptb &                                     Price-to-Book  \\
cr &                             Short Term Leverage  &           quick &                                       Quick Ratio  \\
current &                                   Current Ratio  &          quickg &                                Quick Ratio Growth  \\
currentg &                            Current Ratio Growth  &            rd\_p &     Research and Development scaled by Market Cap  \\
dblock\_n &                                    Blockholders  &         rd\_sale &  Research and Development Expense Scaled by  Sales \\
dbreadth &                 Institutional Change in Breadth  &             rdg &                   Research and Development Growth  \\
deftax &                                  Deferred Taxes  &          revsup &                                  Revenue Surprise  \\
deprate &                               Depreciation Rate  &             roa &                                 Returns on Assets  \\
deprateg &                        Depreciation Rate Growth  &            roag &                       Growth in Returns on Assets  \\
dhhi &                               Ownership Breadth  &          roavol &                                    ROA Volatility  \\
dior &                             Institutional Flows  &             roe &                                 Returns on Equity  \\
dispersion &                     Analyst Forecast Dispersion  &           rv\_iv &                                    Excess Implied  \\
div\_p &                                  Cash Dividends  &          sale\_p &                                    Sales to Price  \\
divinc\_prob &                Probability of dividend increase  &           saleg &                                      Sales Growth  \\
divyield &                                  Dividend Yield  &             scr &                         Short Squeeze Probability  \\
dnoa &                  Change in Net Operating Assets  &     seasonality &                                       Seasonality  \\
dr &                              Long Term Leverage  &        secudebt &                            Secured Long-Term Debt  \\
drg &                              Leverage Expansion  &              sg &                             Abnormal Sales Growth  \\
ev\_ebitda &                      Enterprise Value to EBITDA  &            sgag &  Growth in Selling and General Expenses to Sales   \\
fcf &                       Free Cash Flow Efficiency  &             sio &                   Short Interest Scaled by Supply  \\
fcf\_p &                                 Free Cash Flows  &             sir &                              Short Interest Ratio  \\
filingsize &                    Filing Length and Complexity  &        skew\_atm &                       ATM Put Volatility Skewness  \\
firmtang &               Debt Capacity to Firm Tangibility  &        skew\_otm &                       OTM Put Volatility Skewness  \\
fog\_complex &         Change in Gunning-Fog Readability Index  &           smirk &                     Abnormal Put Volatility Smirk  \\
ft\_mstrong &  Loughran-McDonald modal strong word proportion  &    smog\_complex &                  Change in Smog Readability Index  \\
gprof &                             Gross Profitability  &    specdiv\_prob &                   Probability of special dividend  \\
gprofg &                             Gross Profitability  &         str\_mod &                      Modified Short-term Reversal  \\
hire &                         Growth in Employee Base  &             sue &                                 Earnings Surprise  \\
invg &                    Change in Inventory Turnover  &          taxexp &                            Changes in Tax Expense  \\
invturn &                              Inventory Turnover  &          taxinc &                                    Taxable Income  \\
invturng &                       Inventory Turnover Growth  &       E1          &                 Pollution prevention score    \\
E2              &        Environmental transparency score &  E3              &       Resource efficiency score    \\
CIT1            &           Community \& charity score  & CIT2            &       Human rights score           \\
CIT3            &       Sustainability integration score    & EMP1            &       Compensation \& satisfaction score \\
EMP2      &           Diversity \& rights score        & EMP3            &       Education \& work condition  score    \\
G1    &          Board effectiveness score        & G2              &       Management ethics score    \\
G3              &             Disclosure \& accountability score  & E               &      Environmental score         \\
S               &    Social score  & G               &        Governance score         \\
ESG             &      Averaged E, S, and G scores  & ArticleVolumeTTM               &      Number of articles tagged to SASB categories during the past 12 months    \\
Insight               &    EWMA of Day-to-day variation in response to news  &  OWL Momentum       &      Slope of Insight score \\

\bottomrule
\end{longtable}
\end{landscape}

\bibliography{Paper_ESG.bib}{}
\bibliographystyle{apalike}

\end{document}